\documentclass[11pt,leqno]{article}
\usepackage{amssymb,bbm,amsmath,amsthm,amscd}
\usepackage{vmargin}
\usepackage[dvipsnames]{xcolor}
\usepackage{array}
 
 \usepackage{graphicx}
 \usepackage{subfigure}
\DeclareGraphicsExtensions{.pdf,.eps}

\catcode`\@=11
\newbox\bz@
\newdimen\bdimz@
\def\linethrough#1{\setbox\bz@=\hbox{#1}%
\bdimz@=\ht\bz@ \divide\bdimz@ by 5 \advance\bdimz@ by -\dp\bz@ \ht\bz@=\bdimz@
\leavevmode\hbox{$\overline{\overline{\box\bz@}}$\relax}}
\catcode`\@=12

\def\nbZ{{\mathchoice {\hbox{$\sf\textstyle Z\kern-0.4em Z$}}
{\hbox{$\sf\textstyle Z\kern-0.4em Z$}}
{\hbox{$\sf\scriptstyle Z\kern-0.3em Z$}}
{\hbox{$\sf\scriptscriptstyle Z\kern-0.2em Z$}}}}

\newcounter{compteur}[subsection]

\usepackage[colorlinks=true]{hyperref}

\newtheorem{theorem}{Theorem}[section]

\newtheorem{definition}[theorem]{Definition}

\newtheorem{remark}[theorem]{Remark}

\usepackage{macros}

\makeatletter

\@addtoreset{equation}{section}

\begin{document}

\author{Mejdi Aza\"iez \thanks{Institut Polytechnique de Bordeaux, Laboratoire I2M CNRS UMR5295, France {\tt azaiez@enscbp.fr}},
Tom\'as Chac\'on Rebollo \thanks{Departamento EDAN \& IMUS, Universidad de Sevilla, Spain {\tt chacon@us.es}},
Samuele Rubino \thanks{Departamento EDAN \& IMUS, Universidad de Sevilla, Spain {\tt samuele@us.es}}}
\title{A cure for instabilities due to advection-dominance in POD solution to advection-diffusion-reaction equations.}

\maketitle

\begin{abstract} In this paper, we propose to improve the stabilized POD-ROM introduced by S. Rubino in \cite{Rubino18} to deal with the numerical simulation of advection-dominated advection-diffusion-reaction equations. In particular, we introduce a stabilizing post-processing strategy that will be very useful when considering very low diffusion coefficients, i.e. in the strongly advection-dominated regime. This strategy is applied both for the offline phase, to produce the snapshots, and the reduced order method to simulate the new solutions. The new process of a posteriori stabilization is detailed  in a general framework and applied to advection-diffusion-reaction problems. Numerical studies are performed to discuss the accuracy and performance of the new method in handling strongly advection-dominated cases. 
\end{abstract}


\section{Introduction}\label{sec:Intro}
Reduced-Order Models (ROM) applied to numerical design in modern engineering are a tool that is wide-spreading in the scientific community in the recent years in order to solve complex realistic 
multi-parameters, multi-physics and multi-scale problems, where classical methods such as Finite Difference (FD), Finite Element (FE) or Finite Volume (FV) methods would require up to billions of 
unknowns. On the contrary, ROM are based on a sharp offline/online strategy, and the latter requires a reduced number of unknowns, which allows to face control, optimization, prediction and data 
analysis problems 
in almost real-time, that is, ultimately, a major goal for industrials. The reduced order modeling offline strategy relies on proper choices for data sampling and
construction of the reduced basis (cf. \cite{Hesthaven16}), which will be used then in the online phase, where a proper choice of the reduced model describing the dynamic of the system is
needed. The key feature of ROM is their capability to highly speedup computations, and thus drastically reduce the computational cost of numerical simulations, without compromising too much the
physical accuracy of the solution from the engineering point of view.

\medskip

Among the most popular ROM approaches, Proper Orthogonal Decomposition (POD) strategy provides optimal (from the energetic point of view) basis or modes to represent the dynamics from a given database 
(snapshots) obtained by a full-order system. Onto these reduced basis, a Galerkin projection of the governing equations can be employed to obtain a low-order dynamical system for the basis coefficients. 
The resulting low-order model is named standard POD-ROM, which thus consists in the projection of high-fidelity (full-order) representations of physical problems onto low-dimensional spaces of solutions, 
with a dramatically reduced dimension. These low-dimensional spaces are capable of capturing the dominant characteristics of the solution, their main advantage being that the computations in the 
low-dimensional space can be done at a reduced computational cost. This has led researchers to apply POD-ROM to a variety of physical and engineering problems, including Computational Fluid Dynamics (CFD) 
problems in order to model the Navier--Stokes Equations (NSE), see e.g. \cite{Baiges13, Gunzburger06, Iollo04, Barone12, Beran03, Wang12}. Once applied to the physical problem of interest, POD-ROM can be 
used to solve engineering problems such as shape optimization \cite{Borggaard10a, Borggaard10b} and flow control \cite{Tadmor08, Bergmann08, Peraire99, Tadmor11}.

\medskip

Although POD-ROM can be very computationally efficient and relatively accurate in some flow configurations, they also present several drawbacks. In this paper, we address one of them, 
namely the numerical instability of a straightforward POD-Galerkin procedure applied to convection-dominated flows. The reason of this issue is that, for model reduction purposes, 
one only keeps few modes that are associated to the large eddies of the flow, which should be sufficient to give a good representation of the kinetic energy of the flow, due to the energetic 
optimality of the POD basis functions. However, the main amount of viscous dissipation takes place in the small eddies represented by basis functions that are not taken into account, and thus the 
leading ROM is not able to dissipate enough energy. So, although the disregarded modes do not contain a significant amount of kinetic energy, they have a significant role in the dynamics of the 
reduced-order system. It is then necessary to close the POD-ROM by modeling the interaction between the computed and the unresolved modes. This problem establishes a parallelism to Large 
Eddy Simulations (LES) \cite{Sagaut06} of turbulent flows, where the effect of the smallest flow structures on the largest ones is modeled. Since these are also in non-linear interactions, 
a proper non-linear efficient and accurate closure model should be proposed also in the POD context, considering that in this context the concepts of energy cascade and locality of energy 
transfer are still valid \cite{Couplet03}. To model the effect of the discarded POD modes, 
various approaches have been proposed, both based on physical insights (cf., e.g., the survey in \cite{Wang12}), or on numerical stabilization techniques 
(cf. \cite{Baiges13, BergmannIollo09, Rozza15, IliescuJohn15, IliescuWang14}). 

\medskip

To address this issue, in \cite{Rubino18} a Streamline Derivative projection-based closure modeling strategy for the numerical stabilization of POD-ROM (SD-POD-ROM) has been introduced.
The proposed model has been numerically analyzed for advection-diffusion-reaction equations in the Finite Element (FE) framework, by mainly deriving the corresponding error estimates. Some preliminary numerical tests has been performed in \cite{Rubino18} for a mode\-rate P\'eclet number, showing the efficiency of the proposed method, as well as the increased accuracy over the standard POD-ROM that discovers its well-known limitations very soon in the numerical settings considered, i.e. for moderately low diffusion coefficients.

\medskip

In this paper, we aim to improve this approach by introducing a stabilizing post-processing (SPP) strategy that will be very useful when considering very low diffusion coefficients, i.e. in the strongly advection-dominated regime. The SPP is applied both for the offline phase, to produce the snapshots, and the ROM to simulate the new solutions. The efficiency of this step is highlighted and explained.
  
\medskip

The rest of the paper is organized as follows: in section \ref{sec:VF}, we briefly describe the POD methodology and introduce the SD-POD-ROM for advection-diffusion-reaction problems. In section \ref{sec:PostStab}, we describe the process of a posteriori stabilization in a general framework and how to apply it to the considered problems. Numerical studies are performed in section \ref{sec:NumStad} to discuss the accuracy and efficiency of our method in handling strongly advection-dominated cases, and also its robustness for long time integrations on periodic systems. Finally, Section \ref{sec:Concl} presents the main conclusions of this work and future research directions.

\section{Streamline derivative projection-based POD-ROM}\label{sec:VF}

In this paper, the proposed a posteriori stabilization is preliminary analyzed and tested for the POD-ROM
numerical approximation of advection-dominated advection-diffusion-reaction problems of the form: 
\BEQ\label{eq:uADR}
\left \{
\begin{array}{rcll}
\partial_{t}u + \bv \cdot\nabla u - \nu \Delta u + g u&=&f & \qmbx{in} \Om\times (0,T),\\
u &=& 0 & \qmbx{on} \Ga\times (0,T),\\
u(\xv,0)&=&u^{0}(\xv) & \qmbx{in} \Om,
\end{array}
\right .
\EEQ
where $\bv$ is the given advective field, $\nu<<1$ the diffusion parameter, $g$ the reaction coefficient, $f$ the forcing term, $\Om$ the computational domain in $\mathbb{R}^{d}$, $d=2$ or $3$, $t\in [0,T]$, with $T$ the final time, and $u^{0}$ 
the initial condition. For the sake of simplicity, we have imposed homogeneous Dirichlet boundary conditions on the whole boundary $\Ga=\partial\Om$.

\medskip

To define the weak formulation of problem (\ref{eq:uADR}), let us consider 
the space:
$$
X={H}_{0}^{1} =\left\{v \in H^{1}(\Om)  : v=0  \text{ on } \Ga \right\}, 
$$
where $H^{1}$ is the usual Sobolev space \cite{Brezis11}. 

We shall consider the following variational formulation of (\ref{eq:uADR}):

\medskip
\hspace{1cm} {\em Find $\uv: (0,T)\longrightarrow\Xv$  such that}
\BEQ\label{eq:fvuADR} 
\begin{array}{rcll}
\disp\frac{d}{dt}(u,v) + (\bv \cdot \nabla u, v) + \nu (\nabla u,\nabla v) + g (u,v)  &=&(f,  v) & \forall v\in X,
\end{array} 
\EEQ
where $( \cdot,\cdot )$ stands for the $L^2$-inner product in $\Om$.

\medskip

In order to give a FE approximation of \eqref{eq:fvuADR}, let $\{{\cal T}_{h}\}_{h>0}$ be a family of affine-equivalent, conforming (i.e., without hanging nodes) and regular 
triangulations of $\overline{\Om}$, formed by triang\-les or quadrilaterals ($d=2$), tetrahedra or hexahedra ($d=3$). For any mesh cell $K \in {\cal T}_{h}$,
its diameter will be denoted by $h_K$ and $h = \max_{K \in {\cal T}_{h}} h_K$. We consider $X^{h} \subset X$ a suitable FE space. 
The FE approximation of \eqref{eq:fvuADR} can be written as follows:
\medskip

\hspace{1cm}{\em Find $u_h \in X^{h}$ such that}
\BEQ\label{eq:FEapprox}
\disp\frac{d}{dt}(u_h,v_h) + (\bv \cdot \nabla u_h, v_h) + \nu (\nabla u_h,\nabla v_h) + g (u_h,v_h) =(f,v_h)\quad \forall v_h\in X^{h}.
\EEQ

\medskip

It is well-known that, in the case of low diffusion coefficient $\nu<<1$, the standard Galerkin method \eqref{eq:FEapprox} is generally unstable and leads to globally polluted solutions presenting strong spurious oscillations. In this paper, we thus propose to first consider an offline stabilization procedure, which becomes necessary to deal with the numerical instabilities of the Galerkin method and to generate the snapshots for the online phase with a reasonable accuracy. In particular, we consider a simplification of the Streamline Derivative-based (SD-based) approach used by 
Knobloch and Lube (see \cite{KnoblochLube09}) in the Finite Element (FE) context, which only acts on the high frequencies of the advective derivative. 
This approach consists in adding a filtered advection stabilization term by basically following the streamlines to prevent spurious instabilities due to dominant advection, but using a simple 
interpolation operator on a continuous buffer FE space instead of a local projection operator on a discontinuous enriched FE space (see \cite{IMAJNA17} for more details). 
This stabilization term acts on the high frequencies component (main responsible for numerical oscillations) of the advection/streamline derivative, which seems to be a natural choice when 
dealing especially with strongly advection-dominated configurations. This method falls into the class of Local Projection Stabilization (LPS) methods (cf. \cite{ARCME, Rubino19}). 

\medskip

To briefly recall this approach, assume that the discrete space $X^{h}$ is formed by piecewise polynomial functions of degree $m\geq 2$, e.g. $X^{h}=P_{m}\cap X$, where $P_{m}$ denotes the space of continuous functions whose restriction to each mesh cell $K \in {\cal T}_{h}$ is the Lagrange polynomial of degree less than or equal to $m$. 
We define the scalar product:
$$
(\cdot,\cdot)_{\tau}:L^{2}(\Om)\times L^{2}(\Om) \to \mathbb{R},\quad
(v,w)_{\tau} = \sum_{K\in{\cal T}_{h}}\tau_{K}(v,w)_{K},
$$
and its associated norm:
$$
\nor{v}{\tau}=(v,v)_{\tau}^{1/2},
$$
where for any $K\in{\cal T}_{h}$, $\tau_{K}$ is in general a positive local stabilization parameter (see formula (44) in \cite{Rubino18} for the working expression used in this context).

\medskip

The LPS method by interpolation applied to advection-diffusion-reaction equations is stated by:
\medskip

\hspace{1cm}{\em Find $u_h \in X^{h}$ such that}
\BEQ\label{eq:LPSapprox}
\left \{
\begin{array}{lll}
&&\disp\frac{d}{dt}(u_h,v_h) + (\bv \cdot \nabla u_h, v_h) + (\pi_{h}^{\prime}(\bv \cdot \nabla u_h),\pi_{h}^{\prime}(\bv \cdot \nabla v_h))_{\tau}\\ \\
&+& \nu (\nabla u_h,\nabla v_h) + g (u_h,v_h) =(f,v_h)\quad \forall v_h\in X^{h},
\end{array}
\right .
\EEQ
where $\pi_{h}^{\prime}=Id-\pi_{h}$ is the ``fluctuation operator'', being $Id$ the identity operator, and $\pi_h$ a locally stable interpolation operator from $L^2(\Om)$ onto a projection space $D_h$ defined on the same mesh ${\cal T}_{h}$ and formed by continuous FE (e.g, $D_h=P_{m-1}$), satisfying optimal error estimates (cf. \cite{Chacon13}). In practical implementations, we choose $\pi_h$ as a Scott--Zhang-like \cite{ScottZhang90} linear interpolation operator in the space $P_1$ (since we consider $P_2$ as FE solution space), implemented in the software FreeFem++ \cite{Hecht12}. This interpolant may be defined as: 
$$
\forall x\in \Om,\quad \pi_{h}(v)(x)=\sum_{a\in{\cal N}}{\cal I}_{h}(v)(a)\psi_{a}(x),
$$
where ${\cal N}$ is the set of Lagrange interpolation nodes of $P_1$, $\psi_{a}$ are the Lagrange basis functions associated to ${\cal N}$, and ${\cal I}_{h}$ is the interpolation operator by local averaging of Scott--Zhang kind, which coincides with the standard nodal
Lagrange interpolant when acting on continuous functions (cf. \cite{Chacon13}, section 4).

\subsection{Proper orthogonal decomposition reduced order model}\label{sec:POD-ROM}
For the report to be self-contained, this section briefly presents the computation of a basis for ROM with POD. For more details, the reader is referred to \cite{Chapelle12, Holmes96, Singler14, Sirovich87, Volkwein11}.

\medskip

We first present the continuous version of POD method. Consider a function $u(\xv,t):\Om\times[0,T]\to \mathbb{R}$, and $r\in\mathbb{N}$. Then, the goal of POD consists in finding the set of orthonormal POD basis $\left\{\varphi_{1},\ldots,\varphi_{r}\right\}$ that deliver the best approximation:
\BEQ\label{eq:PODmethCont}
\min\left\| u(\xv,t) - \sum_{i=1}^{r}\left(u(\xv,t),\varphi_{i}\right)_{\mathcal{H}}\varphi_{i} \right\|_{L^2(0,T;\mathcal{H})}^2,
\EEQ
in a real Hilbert space $\mathcal{H}$. Although $\mathcal{H}$ can be any real Hilbert space, in what follows we consider $\mathcal{H}=L^{2}(\Om)$, with induced norm $\nor{\cdot}{}=(\cdot,\cdot)^{1/2}=\left(\disp\int_{\Om}|\cdot|^{2}\right)^{1/2}$.

\medskip 

In the framework of the numerical solution of Partial Differential Equations (PDE), $u$ is usually given at a finite number of times $t_{0},\ldots,t_{N}$, the so-called {\em snapshots}. Let us consider an ensemble of snapshots $\chi=\text{span}\left\{u(\cdot,t_{0}),\ldots,u(\cdot,t_{N})\right\}$, which is a collection of data from either numerical simulation results or experimental observations at time $t_{n}=n\Delta t$, $n=0,1,\ldots,N$ and $\Delta t = T/N$. Then, usually an approximation of the error in the square of the $L^2(0,T)$ norm is considered, e.g., by a modification of the composite trapezoidal rule. Thus, in its discrete version (method of snapshots), the POD method seeks a low-dimensional basis $\left\{\varphi_{1},\ldots,\varphi_{r}\right\}$ that optimally approximates the snapshots in the following sense, see for instance \cite{KunischVolkwein01}:
\BEQ\label{eq:PODmeth}
\min\frac{1}{N+1}\sum_{n=0}^{N}\left\| u(\cdot,t_{n}) - \sum_{i=1}^{r}\left(u(\cdot,t_{n}),\varphi_{i}\right)\varphi_{i} \right\|^2,
\EEQ
subject to the condition $\left(\varphi_{j},\varphi_{i}\right)=\delta_{ij}$, $1\leq i,j \leq r$, where $\delta_{ij}$ is the Kronecker delta.
To solve the optimization problem \eqref{eq:PODmeth}, one can consider the eigenvalue problem:
\BEQ\label{eq:eigen}
K\zv_{i}=\lambda_i\zv_{i},\text{ for } 1,\ldots,r,
\EEQ 
where $K\in \mathbb{R}^{(N+1)\times(N+1)}$ is the snapshots correlation matrix with entries: 
\BEQ\label{CorrM}
K_{mn}=\frac{1}{N+1}\left(u(\cdot,t_{n}),u(\cdot,t_{m})\right),\text{ for } m,n=0,\ldots,N,
\EEQ
$\zv_{i}$ is the $i$-th eigenvector, and $\lambda_{i}$ is the associated eigenvalue. The eigenvalues are positive and sorted in descending order $\lambda_{1}\geq\ldots\geq\lambda_{r}>0$. It can be shown that the solution of \eqref{eq:PODmeth}, i.e. the POD basis, is given by:
\BEQ\label{eq:PODbasis}
\varphi_{i}(\cdot)=\frac{1}{\sqrt{\lambda_i}}\frac{1}{\sqrt{N+1}}\sum_{n=0}^{N}(\zv_{i})_{n}u(\cdot,t_{n}),\quad 1\leq i\leq r,
\EEQ
where $(\zv_{i})_{n}$ is the $n$-th component of the eigenvector $\zv_i$. It can also be shown that the following POD error formula holds \cite{Holmes96, KunischVolkwein01}:
\BEQ\label{eq:PODerr}
\frac{1}{N+1}\sum_{n=0}^{N}\left\| u(\cdot,t_{n}) - \sum_{i=1}^{r}\left(u(\cdot,t_{n}),\varphi_{i}\right)\varphi_{i} \right\|^{2} = \sum_{i=r+1}^{M}\lambda_{i},
\EEQ
where $M$ is the rank of $\chi$. 

\medskip

We consider the following space for the POD setting:
$$
X^{r}=\text{span}\left\{\varphi_{1},\ldots,\varphi_{r}\right\}.
$$

\begin{remark}
Since, as shown in \eqref{eq:PODbasis}, the POD modes are linear combinations of the snapshots, the POD modes satisfy the boundary conditions in \eqref{eq:uADR}. This is because of the particular choice we have made at the beginning to work with homogeneous Dirichlet boundary conditions. In general, one has to manipulate the snapshots set. This is the case, for instance, of steady-state non-homogeneous Dirichlet boundary conditions, for which is preferable to consider a proper lift in order to generate POD modes for the lifted snapshots, satisfying homogeneous Dirichlet boundary conditions. This would lead to work with centered-trajectory method in the POD-ROM setting \cite{IliescuJohn15}.
\end{remark}

In the form it has been presented so far, POD seems to be only a bivariate data compression
or reduction technique, see e.g. \cite{Vega18}. Indeed, equation \eqref{eq:PODmeth} simply says that the POD basis is the best
possible approximation of order $r$ of the given data set. In order to make POD a
predictive tool, one couples the POD with the Galerkin procedure. This, in turn,
yields a ROM, i.e., a dynamical system that represents the evolution in time of the
Galerkin truncation. Thus, the Galerkin POD-ROM uses both Galerkin truncation and Galerkin projection.
The former yields an approximation of the solution by a linear combination of the truncated POD basis:
\BEQ\label{eq:PODsol}
u(\xv,t)\approx u_{r}(\xv,t)=\sum_{i=1}^{r}a_{i}(t)\varphi_{i}(\xv),
\EEQ
where $\left\{a_{i}(t)\right\}_{i=1}^{r}$
are the sought time-varying coefficients representing the POD-Galerkin trajectories. Note that $r<<\mathcal{N}^{dof}$, where $\mathcal{N}^{dof}$ denotes the number of degrees of freedom (d.o.f.) in a full order simulation (e.g., DNS). Replacing $u$ with $u_{r}$ in \eqref{eq:uADR}, using the Galerkin method, and projecting the resulted equations onto the space $X^{r}$, one obtains the standard POD-ROM: 
\BEQ\label{eq:POD-ROM}
\disp\frac{d}{dt}(u_{r},\varphi_{r}) + (\bv\cdot\nabla u_{r},\varphi_{r}) + \nu (\nabla u_{r},\nabla \varphi_{r}) + (g u_{r},\varphi_{r}) =(f,\varphi_{r})\quad \forall \varphi_{r}\in X^{r}.
\EEQ
Despite its appealing computational efficiency, the standard POD-ROM \eqref{eq:POD-ROM} has ge\-nerally been limited to diffusion-dominated configurations. To overcome this restriction, we draw inspiration from the FE context, where stabilized formulations, such as \eqref{eq:LPSapprox} for instance, have been developed to deal with the numerical instabilities of the Galerkin method in advection-dominated configurations.

\subsection{Streamline derivative projection-based method}\label{sec:LPS-POD-ROM}

\medskip
For ease of reading, we recall hereafter the approach leading to the SD-POD-ROM originally introduced and numerically analyzed in \cite{Rubino18}.
Let us introduce the POD space:
$$
\widehat{X}^{r}=\text{span}\left\{\widehat{\varphi}_{1},\ldots,\widehat{\varphi}_{r}\right\},
$$
where $\widehat{\varphi}_{i}$, $i=1,\ldots,r$, are the POD modes associated to $\widehat{K}$, defined as the snapshots correlation matrix with entries:
\BEQ\label{eq:Khat}
\widehat{K}_{mn}=\frac{1}{N+1}\left(\bv\cdot\nabla u(\cdot,t_{n}),\bv\cdot\nabla u(\cdot,t_{m})\right),\quad \text{for } m,n=0,\ldots,N.
\EEQ
Note that for classical POD modes associated to the standard correlation matrix $K_{mn}$, there already exists a theory on convergence rates and error bounds for POD expansions of parameterized solutions of heat equations, see e.g. \cite{MF15, MFT16, MFTal17}. With co-authors of the referred works, we aim to derive a similar analysis for POD modes associated to the advection correlation matrix $\widehat{K}_{mn}$ defined in \eqref{eq:Khat}.

\medskip

We consider the $L^{2}$-orthogonal projection on $\widehat{X}^{r}$, $P_{r}: L^{2}(\Om)\longrightarrow \widehat{X}^{r}$, defined by:
\BEQ\label{eq:Pr}
(u-P_{r}u,\widehat{\varphi}_{r})=0,\quad \forall\widehat{\varphi}_{r}\in\widehat{X}^{r}.
\EEQ
Let $P_{r}^{\prime}=Id-P_{r}$. We propose the Streamline Derivative projection-based POD-ROM (SD-POD-ROM) for \eqref{eq:uADR}: 
\medskip
\BEQ\label{eq:SD-POD-ROM}
\left \{
\begin{array}{lll}
&&\disp\frac{d}{dt}(u_{r},\varphi_{r}) + (\bv\cdot\nabla u_{r},\varphi_{r}) 
+(P_{r}^{\prime}(\bv\cdot\nabla u_{r}),P_{r}^{\prime}(\bv\cdot\nabla \varphi_{r}))_{\tau}
\\ \\
&+& \nu (\nabla u_{r},\nabla \varphi_{r}) + (g u_{r},\varphi_{r}) =(f,\varphi_{r})\quad \forall \varphi_{r}\in X^{r}.
\end{array}
\right .
\EEQ
We introduce the bilinear form $A(u,v)=(\bv\cdot\nabla u,v) + (P_{r}^{\prime}(\bv\cdot\nabla u),P_{r}^{\prime}(\bv\cdot\nabla v))_{\tau} +\nu(\nabla u,\nabla v) + (g u,v)$. The SD-POD-ROM \eqref{eq:SD-POD-ROM} with a backward Euler time discretization reads:
\BEQ\label{eq:SD-POD-ROMdisc}
\disp\frac{1}{\Delta t}(u_{r}^{n+1}-u_{r}^{n},\varphi_{r}) + A(u_{r}^{n+1},\varphi_{r})=(f^{n+1},\varphi_{r})\quad \forall \varphi_{r}\in X^{r}.
\EEQ
\begin{remark}\label{rm:PS-POD-ROM1}
When $\tau_{K}=0$ for any $K\in {\cal T}_{h}$, the SD-POD-ROM \eqref{eq:SD-POD-ROM} coincides with the standard POD-ROM \eqref{eq:POD-ROM}, 
since no numerical dissipation is introduced. Also, note that in this paper we directly consider the projection over the same number $r$ of POD modes retained for the ROM solution. Indeed, due to the slow convergence of the POD eigenvalues associated to the advection correlation matrix $\widehat{K}_{mn}$ in case of very low diffusion (see section \ref{sec:NumStad}) and the fact that error estimates for the SD-POD-ROM are directly proportional to them (cf. \cite{Rubino18}, Theorem 2.11), this improves results obtained by projecting over a number $R<r$, as initially proposed in \cite{Rubino18}.
\end{remark}

\begin{remark}
Note that the SD-POD-ROM \eqref{eq:SD-POD-ROM} rather differs from the VMS-POD-ROM introduced in \cite{IliescuWang13}. Indeed, in \cite{IliescuWang13}, a gradient-based model for the standard POD-ROM is considered, which adds artificial viscosity by a term of the form:
$$
\alpha(\overline{P}_{R}^{\prime}(\nabla u_{r}),\overline{P}_{R}^{\prime}(\nabla \varphi_{r})),
$$  
$\alpha$ being a constant eddy viscosity coefficient, and $\overline{P}_{R}^{\prime}=Id-\overline{P}_{R}$, with $\overline{P}_{R}$ the $L^2$-orthogonal projection on the POD space defined by $\text{span}\{\nabla \varphi_{1},\ldots,\nabla \varphi_{R}\}$, $R<r$, making it applicable just to $H^1$-POD basis, for which the decay of POD eigenvalues is rather slow in presence of strongly advection-dominated configurations.
On the contrary, in the present work, we are adding an advection stabilization term, by basically following the streamlines, which seems to be a more natural choice when dealing especially with strongly advection-dominated regimes. This clearly differentiate the present work with respect to \cite{IliescuWang13}. 

\medskip

Also, the SD-POD-ROM \eqref{eq:SD-POD-ROM} is different from the SUPG-POD-ROM introduced in \cite{IliescuJohn15}, since the former does not involve the full residual (only a streamline derivative stabilization term is introduced), thus presenting a simpler and cheaper structure for practical implementations such as to perform the numerical analysis, and also uses a projection-stabilized structure, which allows to act only on the high frequencies components of the advective derivative: this guarantees an extra-control on them that prevents high-frequency oscillations without polluting the large scale components of the approximation for advection-dominated problems (cf. \cite{Rubino18}, Lemma 2.7). 
\end{remark}

\section{A posteriori stabilization} \label{sec:PostStab}
\def\ca{{\cal A}}
\def\cas{{\cal A^*}}
\def\xh{x_i}
\def\yh{y_i}
\def\wh{w_i}
\def\vh{v_i}
\def\zh{z_i}
\def\Xh{X_i}
\def\Yh{Y_i}
\def\Zh{Z_i}
To describe the process of a posteriori stabilization in a general framework, let us consider an elliptic variational problem:
\BEQ \label{ellvp}
\mbox{ Find } x \in X \mbox{ such that } b(x,w)=l(w)=\langle f,\, w \rangle,\quad \forall w \in X,
\EEQ
where $X$ is a Hilbert space. The form $b$ is defined on $X \times X$  and $l \in X'$, being $X'$ the topological dual of $X$. Consider a family of sub-spaces of finite dimension of $X$, $\{\Xh\}_{i \in {\cal I}}$, for some set of indices ${\cal I}$.  Let us assume that we solve problem \eqref{ellvp} by the Galerkin method on $\Xh$:
\BEQ \label{galerkin}
\mbox{ Find }\xh \in \Xh \mbox{ such that } b(\xh,\wh)=l(\wh),\quad \forall \wh \in \Xh.
\EEQ
Assume that the space $\Xh$ is decomposed into $\Xh=\Yh\oplus \Zh$, where $\Yh$ and $\Zh$ are subspaces of $\Xh$. Let $\xh=\yh+\zh$ be the unique decomposition that $\xh$ admits with $\yh \in \Yh$ and $\zh \in \Zh$. Problem \eqref{galerkin} may be recast as a variational problem for the only unknown $\yh$, as follows. Denote by $\ca$ the operator from $X$ on $X'$ defined by the form $b$; that is for $v\in X$, $\ca v$ is the element of $X'$ defined by: 
$$
\langle \ca v,\, w \rangle = b(v,w),\quad\forall w \in X.
$$
Denote by ${\cal R}_{i} : X' \mapsto \Zh$ the \lq\lq static condensation\rq\rq operator on $\Zh$  generated by the form $b$, defined for $\varphi \in X'$ by:
$$
b( {\cal R}_{i}(\varphi),\, \wh)  = \langle \varphi, \wh \rangle,\quad\forall \wh \in \Zh.
$$
Let us introduce the \lq\lq condensed" variational formulation to problem \eqref{galerkin}:
\BEQ \label{varcond}
\mbox{Find }\yh \in \Xh \mbox{ such that } b_c(\yh,\vh)=l_c(\vh),\quad \forall \vh \in \Yh,
\EEQ
with 
$$
b_c(y,v)=b(y,v)-b({\cal R}_{i}(\cas v),{\cal R}_{i}(\ca y)),\quad l_c(v)= l(v) - b({\cal R}_{i}(\cas v),{\cal R}_{i}(f)),\quad \forall y, \, v \in X;
$$
where $\cas$ denotes the adjoint of the operator $\ca$. 
\par
We next introduce the following definition:
\begin{definition}
The family of finite-dimensional spaces $\{(\Yh,\Zh)\}_{i \in {\cal I}}$, where ${\cal I}$ is a set if indices, is called to satisfy the saturation property if there exists a constant $\alpha>0$ such that
$$
\|\yh\|_X+\|\zh\|_X \le \alpha \, \|\xh + \yh\|_X,\quad \forall \yh \in \Yh, \, \zh\in  \Zh, \quad \forall i \in {\cal I}.
$$
\end{definition}
The saturation property can be viewed as an inverse triangular inequality. It can be readily proved that this property is equivalent to the existence of some constant $\beta >0$ such that
\BEQ \label{prodesc}
|(\yh,\zh)_X| \le (1-\beta)\, \|\yh\|_X \|\zh\|_X, \quad \forall \yh \in \Yh, \, \zh\in  \Zh;
\EEQ
actually we may take $\displaystyle \beta=\frac{2}{\alpha^2}$. Then, we can interpret the saturation property in the sense that the angle between spaces $\Yh$ and $\Zh$, defined by
$$
\arccos \left (\sup_{\yh \in \Yh\setminus \{0\}, \, \zh\in  \Zh\setminus \{0\}} \frac{(\yh,\zh)_X}{\|\yh\|_X \|\zh\|_X}\right )
$$
 is uniformly bounded from below by a positive angle, with respect to $i \in {\cal I}$.

Then, it holds (cf. \cite{chacdom}):
\begin{theorem} \label{testabap}
Assume that the spaces $\Yh$ and $\Zh$ satisfy $\Yh \cap \Zh=\emptyset$. Then:
\begin{enumerate}
\item Let $\xh=\yh+\zh$ be the unique decomposition that $\xh$ admits with $\yh \in \Yh$ and $\zh \in \Zh$. Then, $\xh$ is the solution of the Galerkin method \eqref{galerkin} if and only if 
$\yh$ is the solution of the condensed variational formulation \eqref{varcond}, and
$
\zh={\cal R}_{i}(l-\ca(\yh)).
$
\item
Assume, in addition, that the family of pairs of spaces $\{(\Yh,\Zh)\}_{h \in {\cal I}}$ satisfies the saturation property. Then, there exists a constant $C>0$ such that
\BEQ \label{boundapost}
\|\yh\|_X+\|\zh\|_X \le C\, \|l\|_{X'},\quad \|c_i\|_X \le C \, \|l\|_{X'},
\EEQ
where $c_i={\cal R}_{i}(\ca(\yh))$.
\end{enumerate}
\end{theorem}
We may take advantage of this result to set up an a posteriori stabilization procedure for the Galerkin solution of steady advection-reaction-diffusion equation. In this case, the framework Hilbert space is $X=H^1_0(\Omega)$. Assume that the space $\Yh$ contains in some sense the large scales (or low frequency) component of the space $\Xh$. For instance, if $\Xh$ is a FE space constructed on a grid of a given diameter, $\Yh$ could be a FE subspace of $\Xh$ constructed on a grid with a larger diameter, or with polynomials of lower degree. Also, if $\Xh$ is a POD space, then $\Yh$ could be a subspace formed by a truncated set of basis functions of low frequency. In both cases, $\Zh$ will be a space containing the small scales (or high frequency) components of the space $\Xh$.

\medskip

In this framework, $c_i$ is a representation on $\Zh$ (by means of the static condensation operator) of the small-scale components of the advection-diffusion-reaction operator $\cal A$ acting on the large-scale component $\yh$ of the solution $\xh$. Due to the second estimate in \eqref{boundapost}, $c_i$ is uniformly bounded in $X$ norm. We interpret this bound as an a posteriori stabilization effect. 

\medskip

The stabilization effect largely depends on the actual choice for spaces $\Yh$ and $\Zh$. For instance, for one-dimensional steady advection-diffusion equations with constant advection velocity, diffusion and forcing term, this choice may be made optimal when $\Xh$ is formed by piecewise affine finite elements, as follows. Assume that the space $\Xh$ is built on a grid of grid size $h$, ${\cal T}_h$. The subspace $\Yh$ is formed by piecewise affine finite elements on a grid with double grid size $2h$, ${\cal T}_{2h}$. Then, there is a {\it unique} subspace $\Zh$ such that the solution $\yh$ of the condensed variational formulation \eqref{varcond} coincides with the exact solution $x$ of problem \eqref{ellvp} at the nodes of the grid ${\cal T}_{2h}$. For some other choices of $\Zh$ there could be, however, an over-diffusive effect that yields a large damping of $\yh$ (cf. \cite{chacdom}). 

\medskip
 
Note that to compute $\yh$ from $\xh$ it is not necessary to build the space $\Yh$. Indeed, it suffices to construct a projection operator $\Pi_i\,:\,\xh \in \Xh \mapsto \yh \in \Yh$. To each actual setting for $\Pi_i$ there corresponds a space $\Zh$, as $\Zh =(Id-\Pi_i)(\Xh)$. For Lagrange finite element spaces, in practice the simplest way to compute $\yh$ is to retain just the degrees of freedom of $\xh$ that correspond to the coarser grid on which $\Yh$ is built. Denote by $\{a_1,a_2,\cdots,a_p\}$ the Lagrange interpolation nodes of $\Yh$, and by $\{\varphi_1,\varphi_2,\cdots,\varphi_p\}$ the associated Lagrange basis functions of $\Yh$. There exist a complementary set of interpolation nodes $\{a_{p+1},a_{p+2},\cdots,a_r\}$ and associated basis functions $\{\varphi_{p+1},\varphi_{p+2},\cdots,\varphi_r\}$ such that $\{\varphi_1,\varphi_2,\cdots,\varphi_r\}$ is a basis of $\Xh$. Then, the operator $\Pi_i$ is defined, for any $\displaystyle \xh=\sum_{k=1}^{r} \alpha_k \, \varphi_k \in \Xh$ as:
\BEQ\label{defpii}
\Pi_i\left (\sum_{k=1}^{r} \alpha_k \, \varphi_k \right )= \sum_{k=1}^{p} \alpha_k \, \varphi_k \in \Yh.
\EEQ
The sub scale space $\Zh$ for this procedure is generated by the complementary basis functions $\{\varphi_{p+1},\varphi_{p+2},\cdots,\varphi_r\}$. In \cite{chacdom}, it is proved that the pairs of spaces $\{(\Yh,\Zh)\}_{h \in {\cal I}}$ indeed satisfy the saturation property. In this case the index $i$ may be identified, as usual, with the diameter of the triangulation $h$.

\medskip

For POD approximations, the procedure is quite similar. The space $\Xh$ is generated by the basis functions $\{\varphi_1,\varphi_2,\cdots,\varphi_r\}$,  then the operator $\Pi_i$ is defined by truncation of the POD series $\displaystyle \xh=\sum_{k=1}^{r} \alpha_k \, \varphi_k \in \Xh$ right by \eqref{defpii}, and again the spaces $\Yh$ and $\Zh$ are respectively spanned by $\{\varphi_1,\varphi_2,\cdots,\varphi_p\}$ and $\{\varphi_{p+1},\varphi_{p+2},\cdots,\varphi_r\}$. Note that when the basis functions are orthogonal in $H^1_0(\Omega)$, then the saturation property trivially holds. In this case, the index $i$ may be identified with the dimension $r$ of the space $\Xh$.

\medskip

In this paper we will apply the a posteriori stabilization procedure in the offline stage, in which $\Xh$ is formed by Finite Elements (FE), and also in the online stage, in which $\Xh$ is a POD space. The bilinear form $b$ and the r.h.s. $l$ that we consider are the ones that appear in the problem satisfied by each iterate in the time stepping procedure \eqref{eq:SD-POD-ROMdisc}, that is
\BEQ\label{bilforapost}
b(u,w)=(u,w) + \Delta t \, A(u,w),\quad l(w)= \Delta t \, (f^{n+1},w)+ (u^{n},w) ,\quad \forall u,\, w \in X,
\EEQ
where $u^n$ is the solution at the preceding time step.

\section{Numerical studies}\label{sec:NumStad}
In this section, we present some numerical experiments to mainly assess accuracy and performance of the combination of the Streamline Derivative projection-based stabilization technique \eqref{eq:SD-POD-ROM} with online stabilizing post-processing strategy. We consider the numerical computation of POD-ROM solutions to strongly advection-dominated advection-diffusion-reaction equations. As mentioned above, while for the Full Order Model (FOM) this strategy consists in interpolating the FOM solution on a coarser mesh (in practice, ${\cal T}_{2h}$), for the ROM the a posteriori stabilization consists in truncating the ROM solution once obtained (for the considered numerical experiments, choosing to truncate at $R=r-10$ seems to give the best balance between accuracy and suppression of spu\-rious oscillations). This leads to a computationally efficient and mathematically founded offline/online algorithm (completely separated), implemented over the standard POD-Galerkin ROM. Actually, two applications (offline and online) of the stabilized post-processing technique are studied in this paper, where we will show the good  performances of this technique to stabilize highly oscillating FOM and ROM numerical solutions of strongly advection-dominated problems. 

\medskip

The first numerical test \ref{subsec:2DRotCyl} concerns an almost pure transient transport problem with a rotating cylinder. The second numerical test \ref{subsec:2DTravWave} concerns a 2D traveling wave displaying a sharp internal layer moving in time.
In both cases, we employ $P_2$ (piecewise quadratic) FE on relatively coarse uniform spatial discretizations, and the backward Euler method for temporal discretization with time step $\Delta t = 10^{-3}$. In particular, FE meshes were significantly coarser than the width of the internal layers, which is common in practice.
The open-source FE software FreeFem++ \cite{Hecht12} has been used to run all numerical experiments.

\subsection{2D Rotating cylinder}\label{subsec:2DRotCyl}
\begin{figure}[htb]
\begin{center}
\centerline{\includegraphics[scale=0.2]{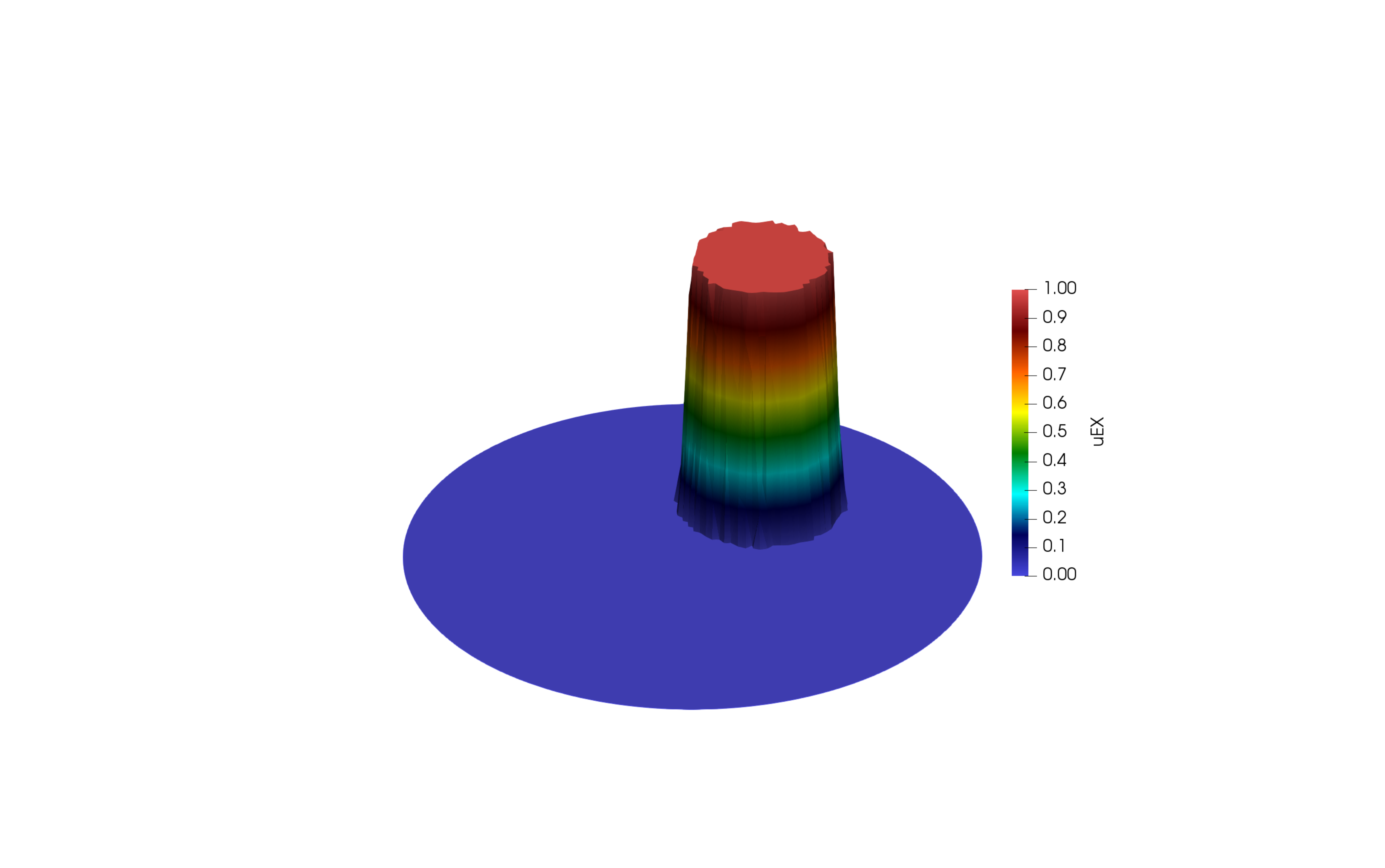}}
\caption{Example \ref{subsec:2DRotCyl}: Initial condition.}\label{fig:01}
\end{center}
\end{figure}

In this section, an almost pure transient transport problem with a rotating body will be considered. In particular, this problem is given in the unit disc $\Om=\{(x,y)\in\mathbb{R}^2 : x^2 + y^2<1\}$ by the advection-diffusion-reaction equation \eqref{eq:uADR} with advection field $\bv=(-y,x)^{\mathrm{T}}$, reaction coefficient $g=0$, forcing term $f=0$, and a very small value for the diffusion parameter $\nu=10^{-20}$, as in \cite{AhmedMatthies16}. The initial condition $u^{0}$ is given by: 
\BEQ\label{ex}
u^0=0.5\left[\tanh\left(\disp\frac{e^{-10[(x-0.3)^2 + (y-0.3)^2 - 0.5]}}{10^{-3}}\right)+1\right],
\EEQ 
which consists in a cylinder of height $1$ centered at $(0.3,0.3)$, as shown in figure \ref{fig:01}. This condition is smooth, but has a sharp layer with thickness of order $10^{-3}$. The mesh is uniform with $256$ triangles along the boundary of $\Om$, which leads to mesh size $h=4.26\cdot 10^{-2}$, thus the layer is under-resolved. The rotation is counter-clockwise and the solution after complete revolutions should be essentially the same as the initial condition, since the diffusion parameter $\nu=10^{-20}$ is very small. A pure transient transport problem with this data was considered in \cite{Burman10}.

This example leads to a {\bf strongly advection-dominated problem}, 
and therefore an offline stabilization procedure becomes necessary to deal with the numerical instabilities of the Galerkin method. 
As announced in section \ref{sec:VF}, in this work we preliminarily consider the LPS-FE by interpolation Method (LPS-FEM) given by \eqref{eq:LPSapprox}, to which we further apply the a posteriori stabilization described in section \ref{sec:PostStab}. 

\subsubsection{Short time behavior}\label{subsubsec:STB}
\begin{figure}[htb]
\begin{center}
\centerline{\includegraphics[scale=0.4]{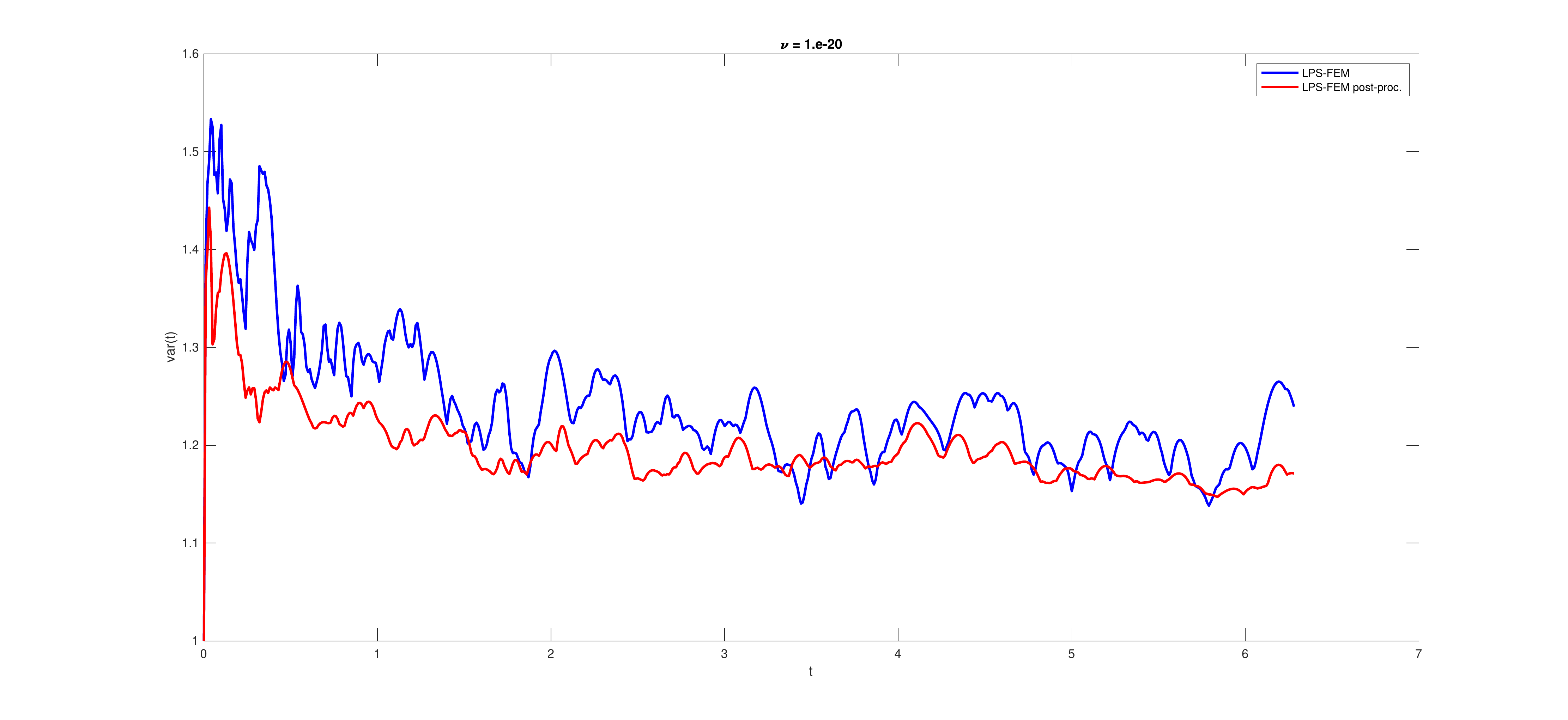}}
\caption{Example \ref{subsubsec:STB}: Measure FOM $var(t)$ for under- and overshoots.}\label{fig:03}
\end{center}
\end{figure}

In first instance, we just compute one complete revolution of the cylinder being transported around the unit disc, i.e. the computational time interval is $[0,T]=[0,2\pi]$.
Note that the application of the a posteriori stabilization described in the previous section further improves the accuracy provided by the LPS-FEM, as shown in figure \ref{fig:03}, where we consider:
$$
var(t)=\max_{(x,y)\in\Om} u_{h}(x,y,t)-\min_{(x,y)\in\Om} u_{h}(x,y,t),
$$
as measure for under- and overshoots, as in \cite{JohnNovo11}. Indeed, we observe that, even if both methods gives similar error levels, LPS-FEM with post-processing is superior to LPS-FEM, for which the quantity $var(t)$ shows much larger oscillations. Note that the optimal value of $var(t)$ equals to $1$ for all $t$.

\begin{figure}[htb]
\begin{center}
\centerline{\includegraphics[scale=0.4]{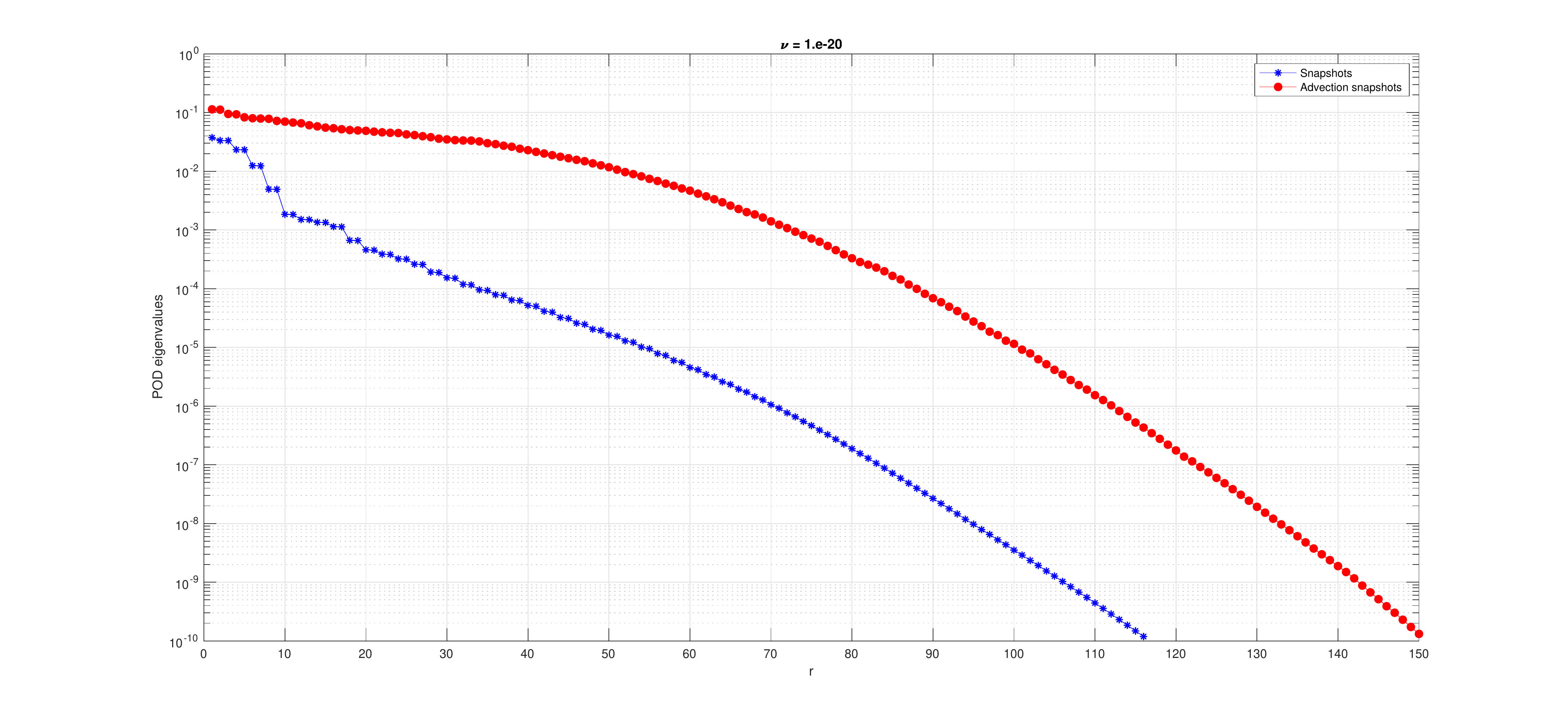}}
\caption{Example \ref{subsubsec:STB}: POD eigenvalues.}\label{fig:04}
\end{center}
\end{figure}

As for the online phase, we perform a comparison between the SD-POD-ROM \eqref{eq:SD-POD-ROM} by considering the application or not of the a posteriori stabilization technique mentioned above, adapted to the POD-ROM framework. Similar results (therefore not reported) are obtained in this case by considering the standard POD-ROM \eqref{eq:POD-ROM}. The POD modes are generated in $L^{2}$ by the method of snapshots by storing every tenth FOM solution in the computational time interval $[0,T]=[0,2\pi]$, so that $629$ snapshots were used. POD basis were constructed by using LPS-FEM with stabilizing post-processing, to limit the influence of POD noisy data in the online phase. In figure \ref{fig:04}, we show the decay of POD eigenvalues associated both to the snapshots correlation matrix \eqref{CorrM} and the advection correlation matrix \eqref{eq:Khat} in this case. Comparing this figure with next figures \ref{fig:2} and \ref{fig:7}, we observe that this test, despite the smaller diffusion coefficient, is smoother with respect to the following test concerning the 2D traveling wave problem, due to the faster decay of POD eigenvalues. 

\begin{figure}[htb]
\begin{center}
\centerline{\includegraphics[scale=0.65]{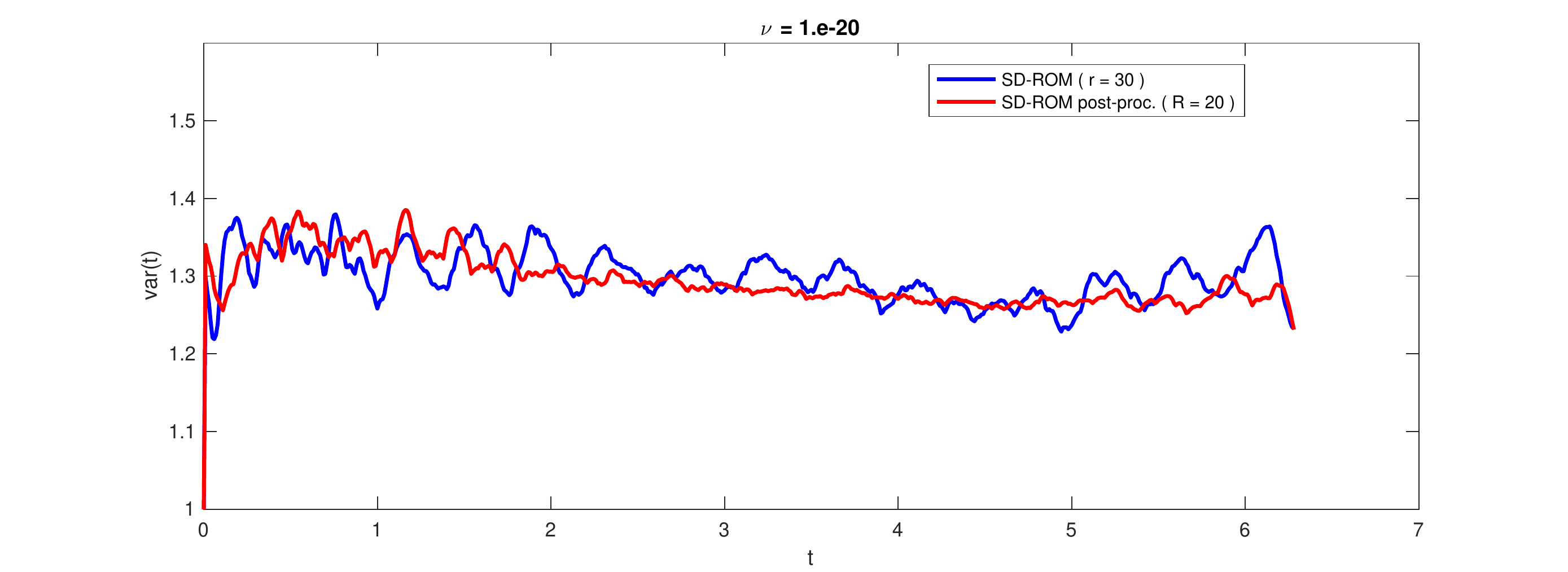}}
\centerline{\includegraphics[scale=0.65]{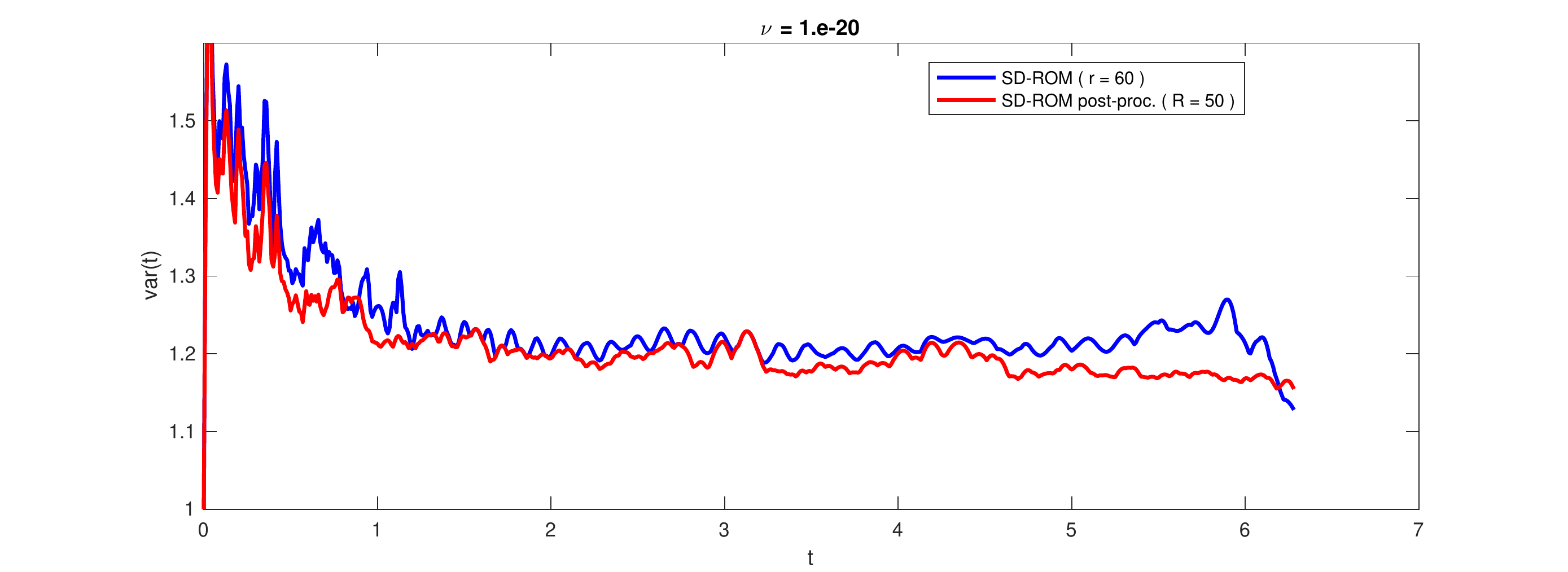}}
\centerline{\includegraphics[scale=0.65]{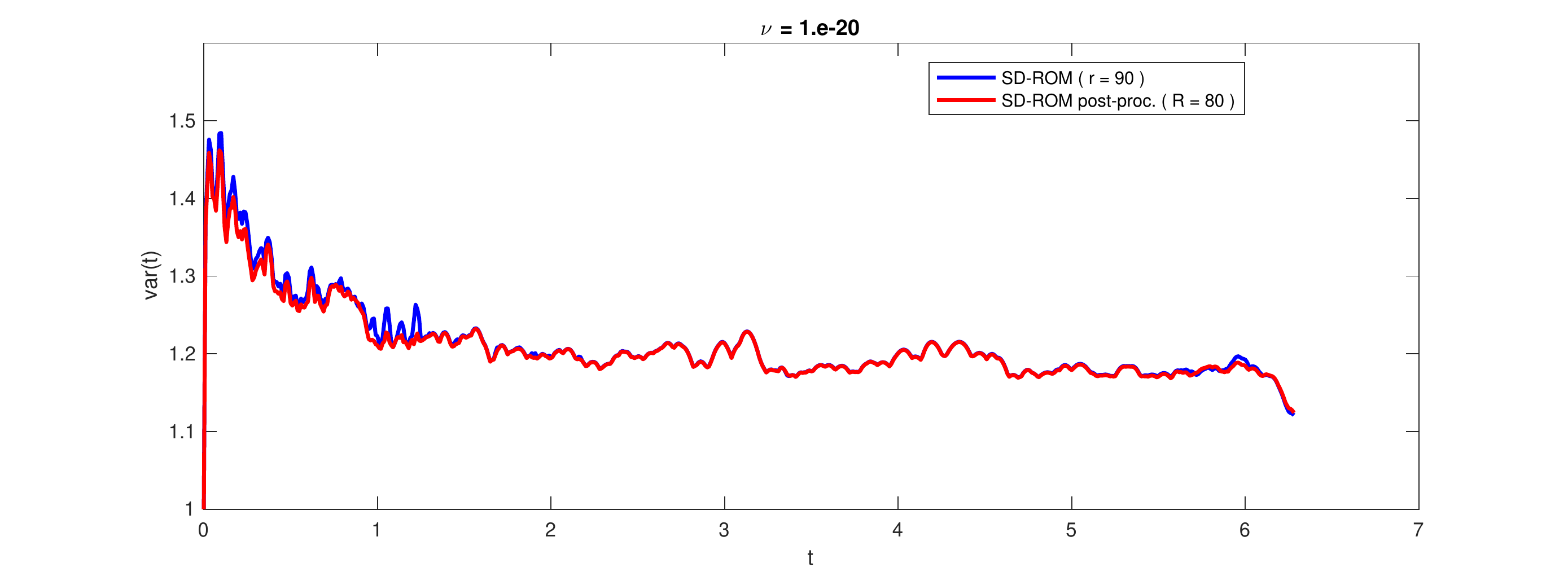}}
\caption{Example \ref{subsubsec:STB}: Measure $var(t)$ for under- and overshoots for different ROM at $r=30, 60, 90$ (from top to bottom).}\label{fig:05}
\end{center}
\end{figure}

\medskip
To check the temporal behavior of the online spurious oscillations, we compute $var(t)$ as:
$$
var(t)=\max_{(x,y)\in\Om} u_{r}(x,y,t)-\min_{(x,y)\in\Om} u_{r}(x,y,t),
$$
for the different ROM, tested in the same computational time interval $[0,T]=[0,2\pi]$ where the snapshots were computed. The corresponding results are displayed in figure \ref{fig:05}, where we evaluate the measure $var(t)$ for under- and overshoots at $r=30, 60, 90$ (from top to bottom) both for SD-POD-ROM (SD-ROM) and SD-POD-ROM with online stabilized post-processing (SD-ROM post-proc.). To compute $var(t)$ for SD-ROM post-proc., note that the online stabilized post-process is applied at the end of each time iteration, although the post-processed solution is not used to continue iterating in time so that this is computationally very cheap. It is interesting to observe that, although the first $r=30$ POD modes already capture more than $99\%$ of the system's kinetic energy (see table \ref{tab:0}), both ROM yield poor quality results for which $var(t)$ oscillates around $1.3$ for all $t$, reflecting the complexity of the problem. Augmenting the number of POD modes causes the decrease of $var(t)$ to values close to $1.1$ after one full turn. Similarly to the offline phase, we observe that, even if both online methods gives similar error levels, SD-ROM with online post-processing is superior to SD-ROM, for which the quantity $var(t)$ shows much larger oscillations. Differences are reduced augmenting the number of POD modes, as expected.

\begin{figure}[htb]
\begin{center}
\includegraphics[scale=0.134]{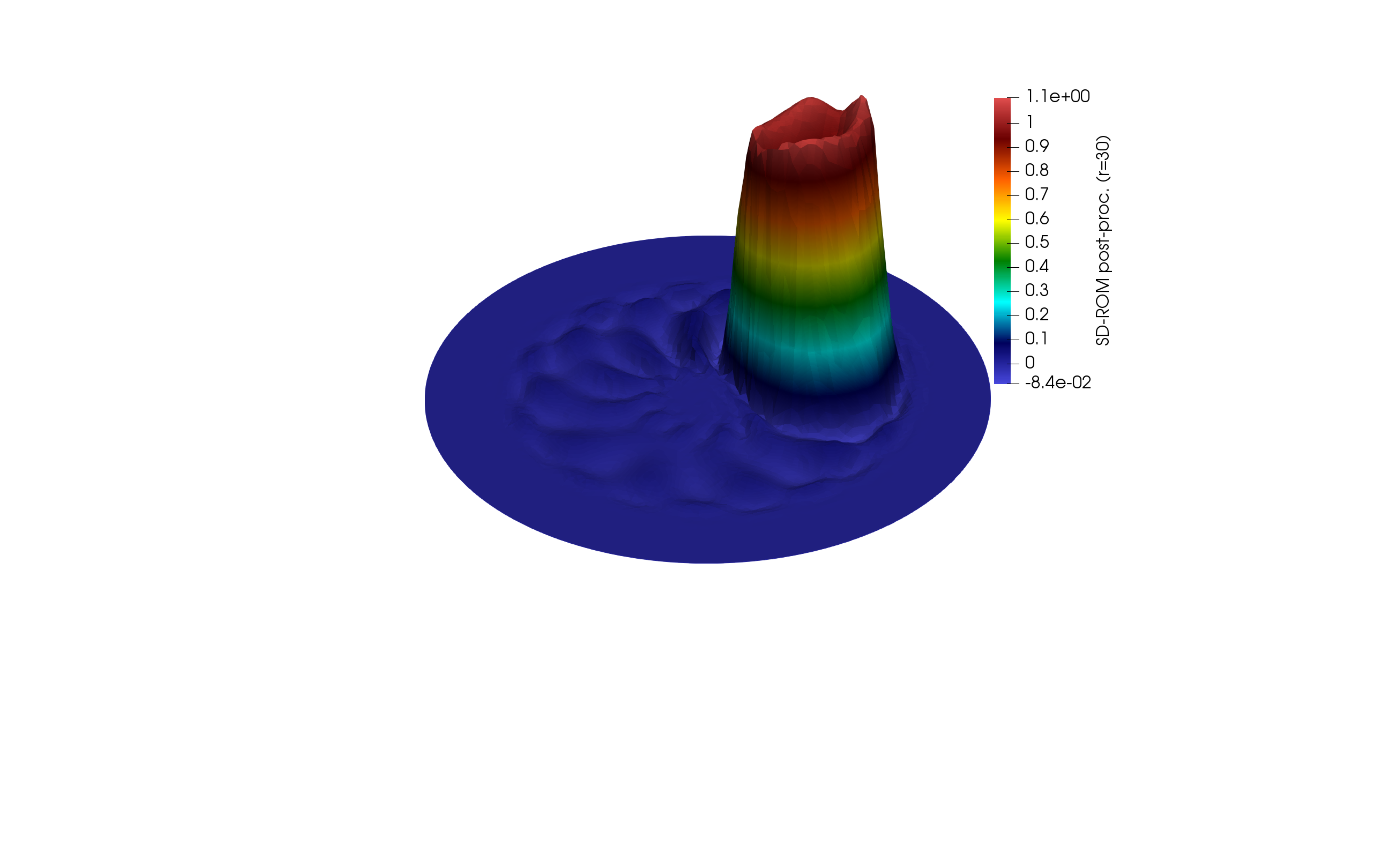}
\includegraphics[scale=0.134]{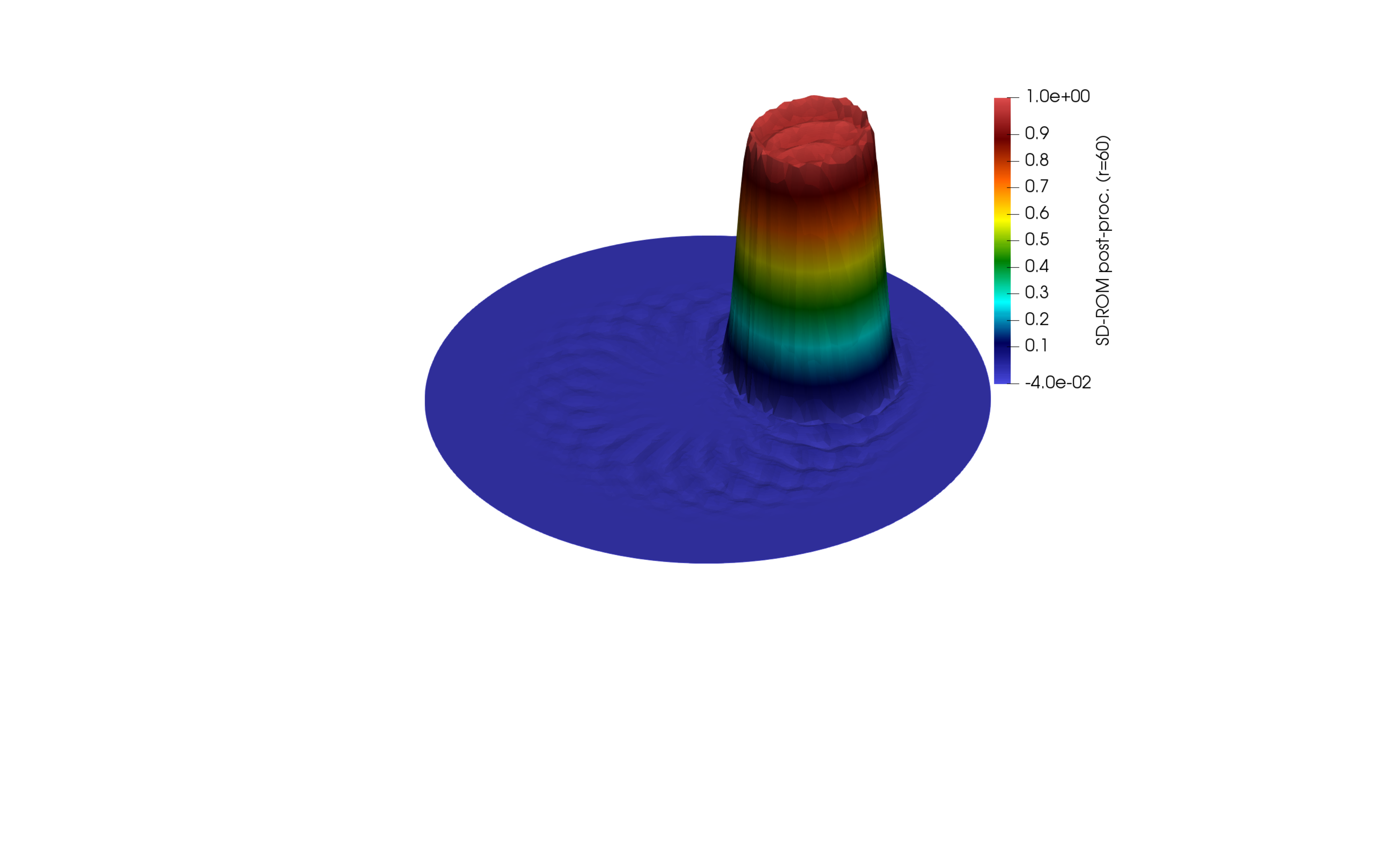}
\includegraphics[scale=0.134]{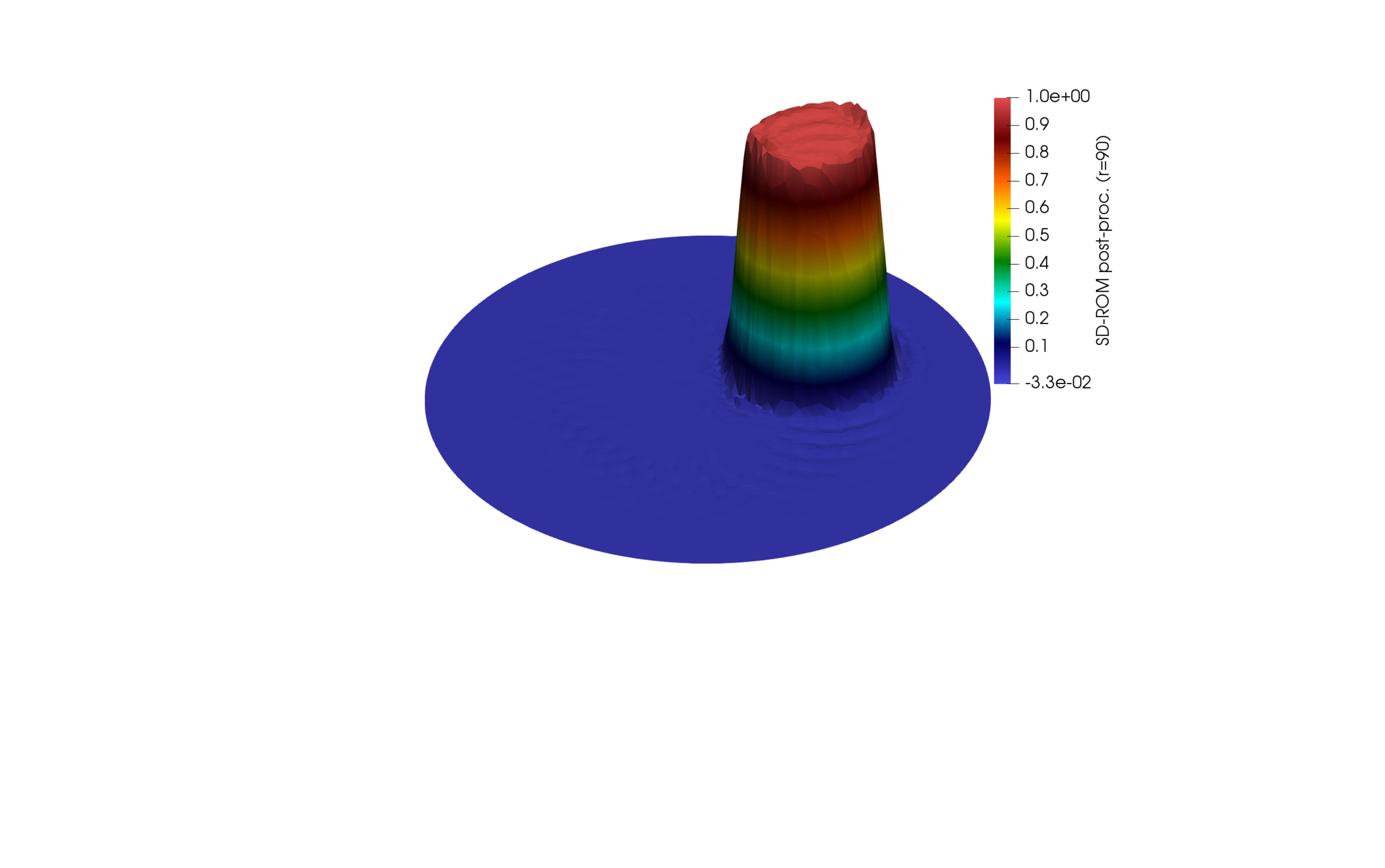}
\caption{Example \ref{subsubsec:STB}: Numerical solution for SD-ROM with online stabilizing post-processing at $T=2\pi$ for $r=30, 60, 90$ (from top to bottom).}\label{fig:06}
\end{center}
\end{figure}

\begin{table}[htb]
$$\hspace{-0.1cm}
\begin{tabular}{|c|c|c|c|}
\hline
$\nu=10^{-20}$ & $r=30$ & $r=60$ & $r=90$\\
\hline
Captured system's $E_{kin} (\%)$ & $99.35$ & $99.99$ & $> 99.99$\\
\hline
\end{tabular}$$\caption{Example \ref{subsubsec:STB}: Captured system's kinetic energy at $r=30, 60, 90$.}\label{tab:0}
\end{table}
\medskip
To give a qualitative comparison, we report in figure \ref{fig:06} the final numerical solutions after one full turn obtained using the best performing SD-ROM with online a posteriori stabilization for $r=30, 60, 90$ (from top to bottom). To compute them, note that the online stabilized post-process only applies to the ROM solutions just at the end, so that this is again computationally very cheap. We observe that numerical unphysical oscillations are gradually reduced by increasing the number of POD modes, allowing to compute a rather accurate final solution.

\subsubsection{Long time beahvior}\label{subsubsec:LTB}
The aim of this section is to check the long time behavior of the spurious oscillations measured by $var(t)$ (cf. \cite{AhmedMatthies16}), and also the performance of the SD-ROM over a larger time interval with respect to the one used to compute the snapshots and generate the POD modes (cf. \cite{StabileRozza18}). This would assess the robustness and prediction/extrapolation ability of the SD-ROM for long time integrations on this almost periodic system.

\medskip

\begin{figure}[htb]
\begin{center}
\centerline{\includegraphics[scale=0.3]{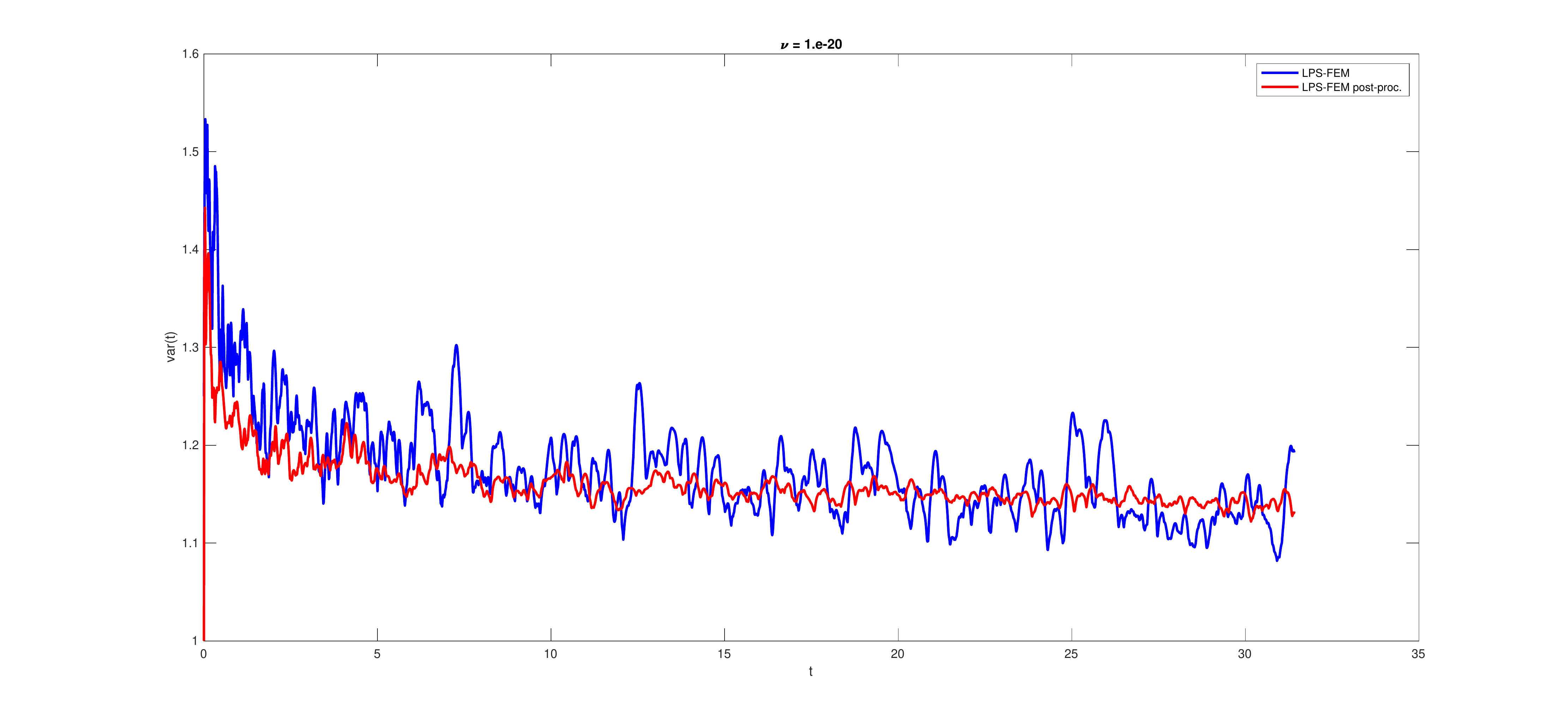}}
\caption{Example \ref{subsubsec:LTB}: Measure FOM $var(t)$ for under- and overshoots.}\label{fig:07}
\end{center}
\end{figure}

To do so, we first compute LPS-FEM with and without post-processing till $T=10\pi$, which corresponds to five complete revolutions. After an initial decreasing phase, the quantity $var(t)$ almost stabilizes in the range $[1.1,1.2]$, see figure \ref{fig:07}. Again, it is interesting to observe that, even if both methods gives similar error levels, the quantity $var(t)$ shows much larger oscillations for LPS-FEM without post-processing. 

\begin{figure}[htb]
\begin{center}
\centerline{\includegraphics[scale=0.3]{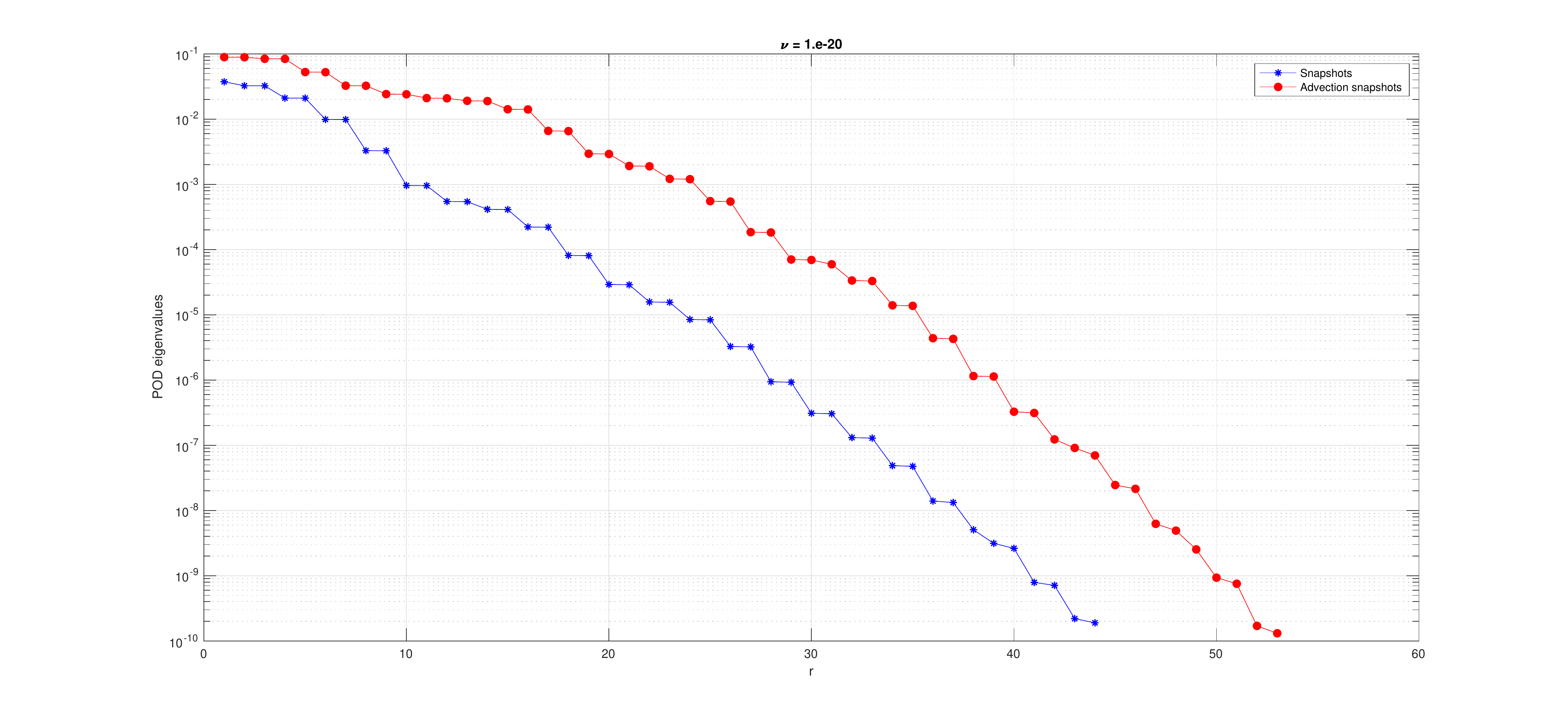}}
\caption{Example \ref{subsubsec:LTB}: POD eigenvalues.}\label{fig:08}
\end{center}
\end{figure}

\medskip
As for the online phase, in this case only the last simulated revolution $[8\pi,10\pi]$ is used to collect the snapshots for the POD basis generation, since we are interested in the correct behavior of the SD-ROM during the almost stable response regime. Within this time range, the POD basis are generated in $L^2$ by the method of snapshots by storing every tenth solution, so that $629$ snapshots were used. POD basis were constructed by using LPS-FEM with stabilizing post-processing, to limit the influence of POD noisy data in the online phase. In figure \ref{fig:08}, we show the decay of POD eigenvalues associated both to the snapshots correlation matrix \eqref{CorrM} and the advection correlation matrix \eqref{eq:Khat} in this case.

\begin{figure}[htb]
\begin{center}
\centerline{\includegraphics[scale=0.3]{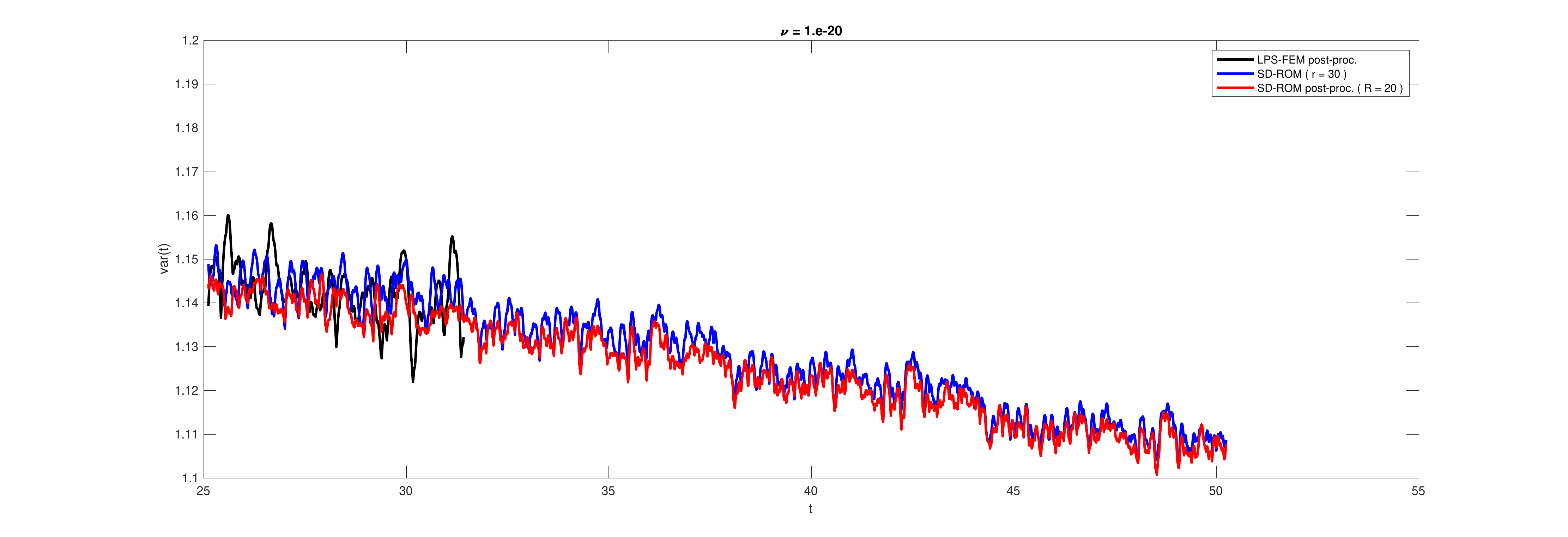}}
\caption{Example \ref{subsubsec:LTB}: Long time behavior of measure $var(t)$ for under- and overshoots for different ROM at $r=30$.}\label{fig:09}
\end{center}
\end{figure}

\medskip
To check the long time behavior of the online spurious oscillations measured by $var(t)$, a comparison between SD-ROM with and without online stabilized post-processing is performed in the time range $[8\pi,16\pi]$, which is four times wider with respect to the time window used for the generation of the POD basis. The corresponding results are displayed in figure \ref{fig:09}, where we evaluate the measure $var(t)$ for under- and overshoots at $r=30$ both for SD-ROM and SD-ROM post-proc. in $[8\pi,16\pi]$, and compare it with the FOM one in the snapshots time range $[8\pi,10\pi]$. Note that for $r=30$ more than $99.99\%$ of the system's kinetic energy is captured in this case. Both SD-ROM gives here almost similar and reliable results for long time integration, being SD-ROM post-proc. slightly superior to SD-ROM, and seems to rightly follow the trend initially given by the FOM by approaching values close to $1.1$.

\subsection{2D Traveling wave}\label{subsec:2DTravWave}
The mathematical model used for the numerical studies in this section is the advection-diffusion-reaction equation \eqref{eq:uADR} with the following parameter choices: computational spatial domain $\Om=(0,1)^2$, 
computational time interval $[0,T]=[0,1]$, advection field $\bv=\left(\cos{\disp\frac{\pi}{3}},\sin{\disp\frac{\pi}{3}}\right)^{\mathrm{T}}$, reaction coefficient $g=1$, and two low values 
for the diffusion parameter: $\nu\in\left\{10^{-6},\,10^{-8}\right\}$ . The forcing term $f$ and initial condition $u^{0}$ are chosen to satisfy the exact solution: 
\BEQ\label{ex}
u(x,y,t)=0.5\sin(\pi x)\sin(\pi y)\left[\tanh\left(\disp\frac{x+y-t-0.5}{4\sqrt{\nu}}\right)+1\right],
\EEQ 
which simulates a 2D traveling wave displaying a sharp internal layer of width $\mathcal{O}(\sqrt{\nu})$ moving in time. 

\medskip

This example leads again to a {\bf strongly advection-dominated problem}, 
and therefore an offline stabilization procedure becomes necessary to deal with the numerical instabilities of the Galerkin method. 
As in the previous section, we preliminarily consider the LPS-FE by interpolation Method (LPS-FEM) given by \eqref{eq:LPSapprox}, to which we further apply the a posteriori stabilization described in section \ref{sec:PostStab}. First, we consider the intermediate case $\nu=10^{-6}$, for which the application or not of the a posteriori stabilization technique described in the previous section almost gives a similar accuracy to compute the snapshots. Then, we consider the limit case $\nu=10^{-8}$, for which instead the application of the a posteriori stabilization further improves the accuracy provided by the LPS-FEM, as we will see in the next sections. 

\medskip

As for the online phase, we perform a comparison between the standard POD-ROM \eqref{eq:POD-ROM} and the SD-POD-ROM \eqref{eq:SD-POD-ROM}, by considering in both cases the application or not of the a posteriori stabilization technique mentioned above, adapted to the POD-ROM framework. The POD modes are generated in $L^{2}$ by the method of snapshots by storing every tenth solution, so that $101$ snapshots were used. Since the forcing term $f$ is time-dependent, the global load vectors are stored for later use in the tested POD-ROM.

\medskip

Besides plots of the computed final ROM solutions with higher accuracy, we also performed a comparison between the different types of studied ROM by evaluating the deviation $e_0$ for the final solution profile along the mean diagonal (connecting vertices $(0,0)$ and $(1,1)$) of the computational domain from the corresponding exact solution profile in a normalized discrete $L^{2}$-norm subject to:
\BEQ\label{errROM}
e_{0}^{ROM}=\left[\frac{\disp\int_{0}^{\sqrt{2}}\left|u_{ex}^{fin}-u_{ROM}^{fin}\right|^2}{\disp\int_{0}^{\sqrt{2}}\left|u_{ex}^{fin}\right|^2}\right]^{1/2},
\EEQ 
with obvious notation. An analogous for the different types of studied FOM has also been computed, by considering:
\BEQ\label{errFOM}
e_{0}^{FOM}=\left[\frac{\disp\int_{0}^{\sqrt{2}}\left|u_{ex}^{fin}-u_{FOM}^{fin}\right|^2}{\disp\int_{0}^{\sqrt{2}}\left|u_{ex}^{fin}\right|^2}\right]^{1/2}.
\EEQ

\subsubsection{Case $\nu = 10^{-6}$}\label{subsubsec:Ex1}
\begin{figure}[htb]
\begin{center}
\centerline{\includegraphics[scale=0.4]{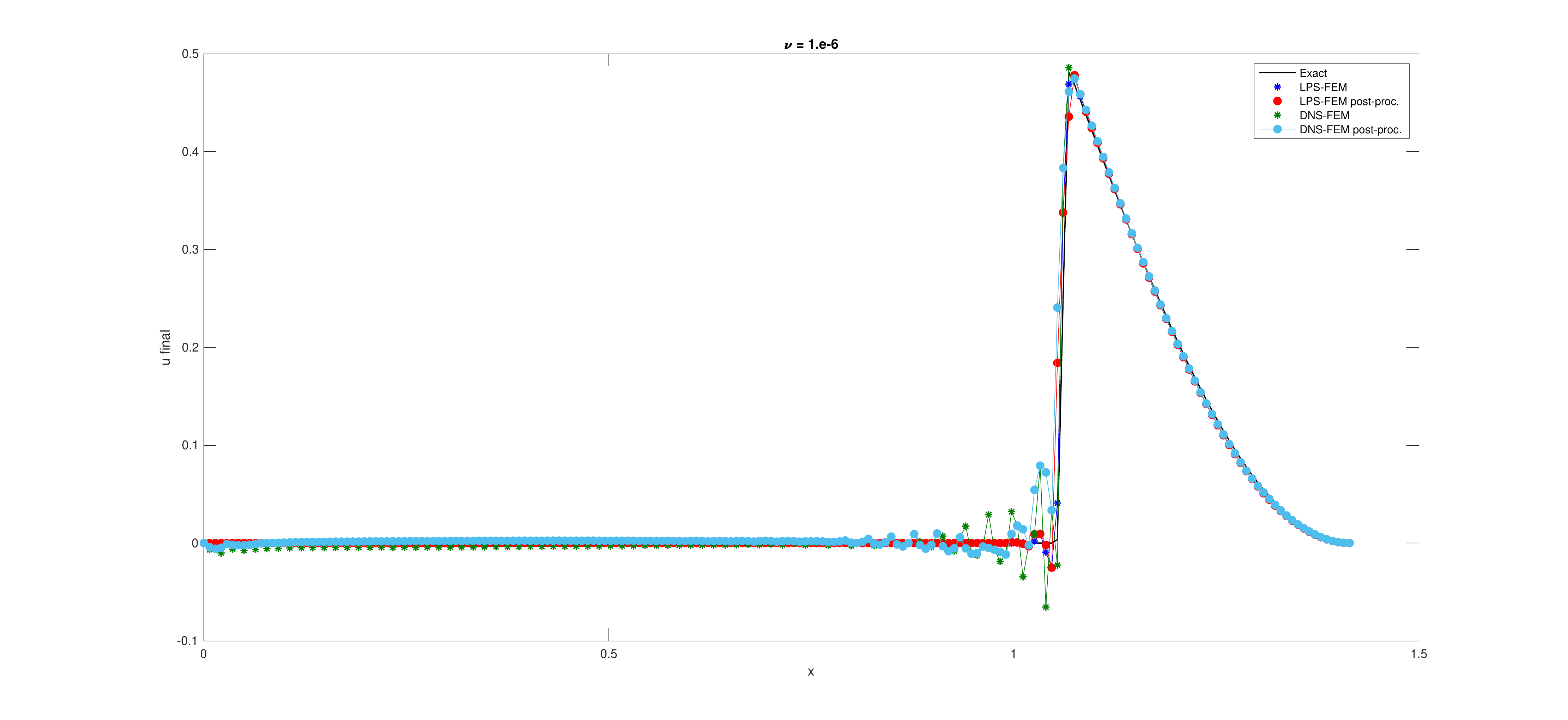}}
\caption{Example \ref{subsubsec:Ex1}: Final solution profiles along the mean diagonal for different FOM.}\label{fig:1}
\end{center}
\end{figure}

In this case, we consider a uniform triangular mesh with mesh size $h=1.41\cdot 10^{-2}$, which is relatively coarse with respect to the width of the internal layer. First, we tested diffe\-rent FOM: the Direct Numerical Simulation \eqref{eq:FEapprox} (DNS-FEM), where no stabilization is introduced, a DNS with stabilized post-processing (DNS-FEM post-proc.), the LPS (by interpolation)-FEM \eqref{eq:LPSapprox} (LPS-FEM), and the LPS-FEM with stabilized post-processing (LPS-FEM post-proc.). In figure \ref{fig:1}, we show for the different methods the final solution profiles along the mean diagonal of the computational domain compared with the corresponding exact solution profile.

\begin{table}[htb]
$$\hspace{-0.1cm}
\begin{tabular}{|c|c|}
\hline
Offline methods & $e_{0}^{FOM}$, $\nu=10^{-6}$\\
\hline
DNS-FEM & $0.1828$\\
\hline
DNS-FEM post-proc. & $0.1257$\\
\hline
LPS-FEM & $0.0576$\\
\hline
LPS-FEM post-proc. & $0.0618$\\
\hline
\end{tabular}$$\caption{Example \ref{subsubsec:Ex1}: $L^2$-norm of the deviation from the final exact solution profile along the mean diagonal for different FOM.}\label{tab:1}
\end{table}
\medskip
From this figure, it is evident that a DNS (i.e., no stabilization) gives oscillatory results, which are only in part corrected by applying the a-posteriori stabilization. Thus, since the problem is advection-dominated and the solution has already a steep internal layer, the use of a stabilized discretization is necessary when using relatively coarse meshes. For this purpose, we considered LPS by interpolation method, for which oscillations are rather reduced, and application or not of the a-posteriori stabilization almost gives similar results. A quantitative comparison between the different FOM is given in table \ref{tab:1}, where the deviation $e_{0}^{FOM}$ from the final exact solution profile along the mean diagonal in a norma\-lized discrete $L^{2}$-norm subject to \eqref{errFOM} is displayed. We may observe that, while for DNS methods errors are greater than $10\%$, for LPS-FEM methods are comparable and below $10\%$, being slightly better for the LPS-FEM method without a posteriori stabilization.

\begin{figure}[htb]
\begin{center}
\centerline{\includegraphics[scale=0.4]{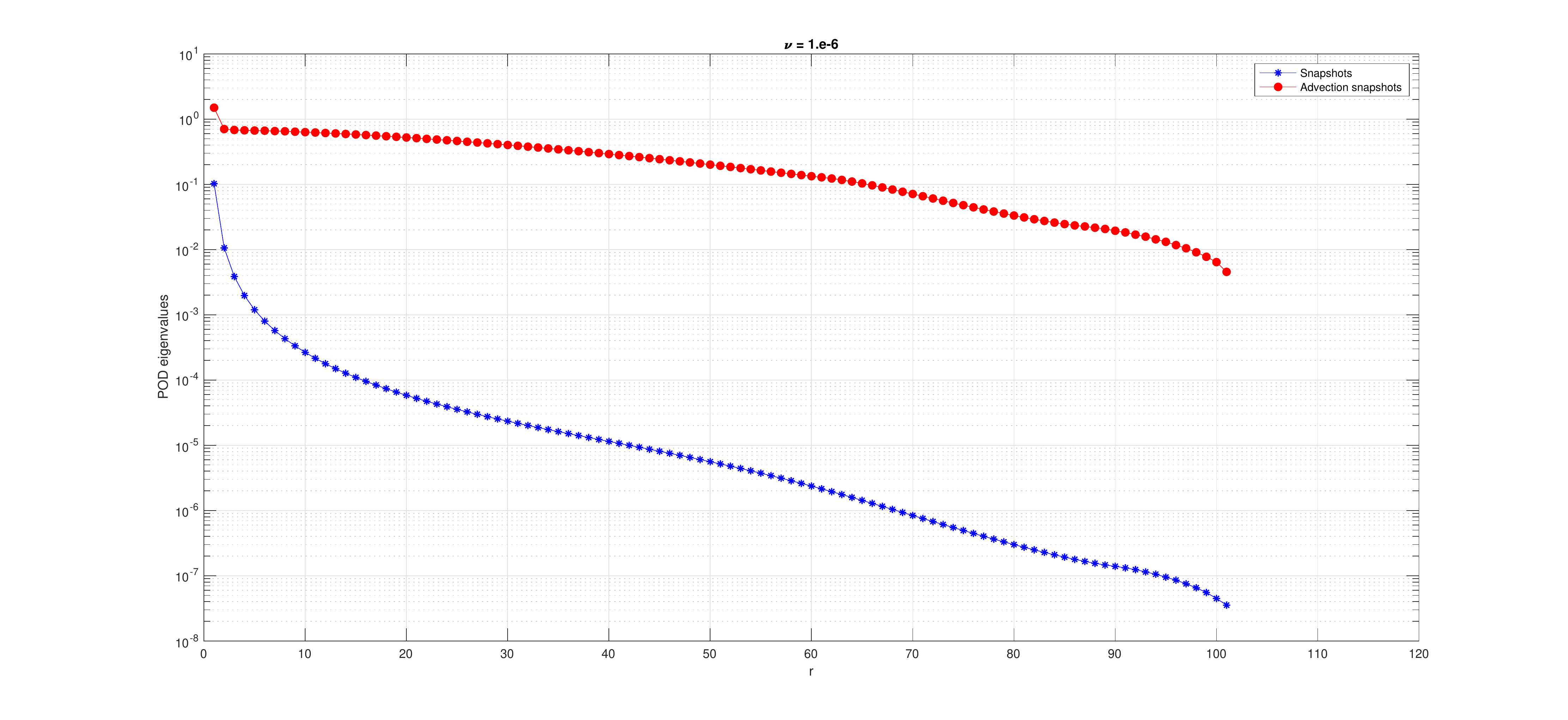}}
\caption{Example \ref{subsubsec:Ex1}: POD eigenvalues.}\label{fig:2}
\end{center}
\end{figure}
\medskip
So, for this case, POD basis were constructed by using LPS-FEM method \eqref{eq:LPSapprox}, and the studied ROM thus used just slightly noisy POD data, which is unavoidable for strongly advection-dominated problems on realistic grids. In figure \ref{fig:2}, we show the decay of POD eigenvalues associated both to the snapshots correlation matrix \eqref{CorrM} and the advection correlation matrix \eqref{eq:Khat}. One can observe that the decay of the POD eigenvalues asso\-ciated to the advection correlation matrix is rather slow, due to the low diffusion. However, adding the corresponding stabilization term in the online phase greatly improves the results over the standard POD-ROM, since allows to control the high frequencies components of the advective derivative, main responsible for numerical oscillations.

\begin{figure}[htb]
\begin{center}
\centerline{\includegraphics[scale=0.35]{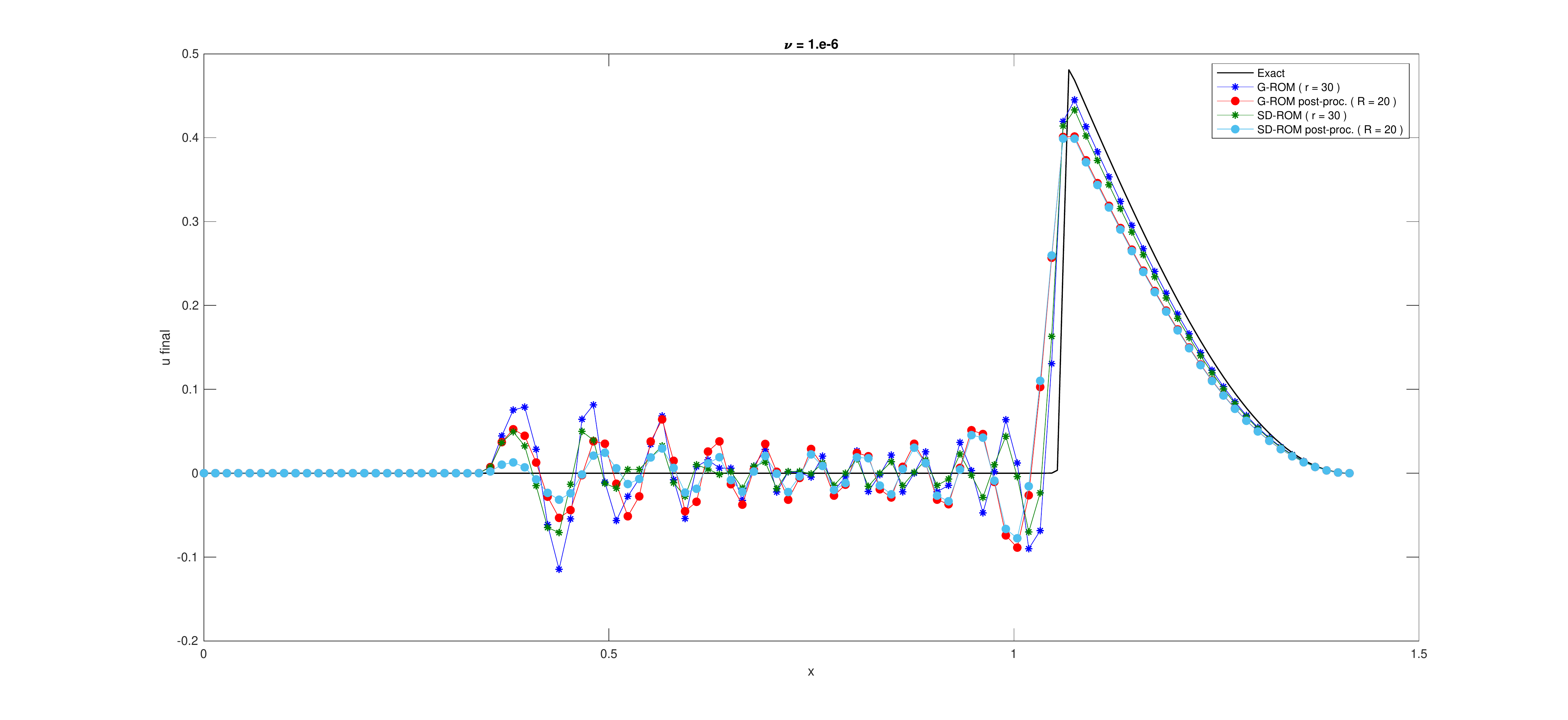}}
\centerline{\includegraphics[scale=0.35]{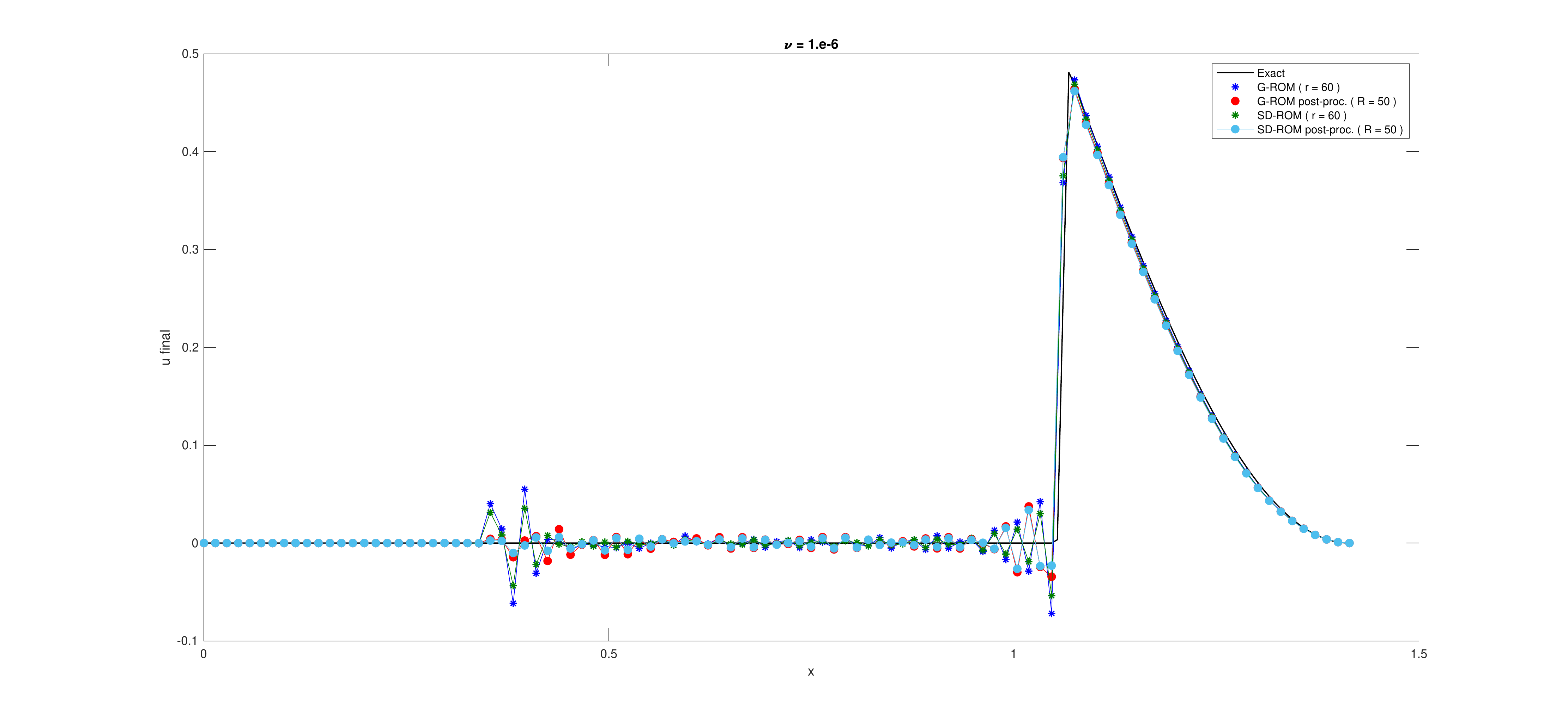}}
\centerline{\includegraphics[scale=0.35]{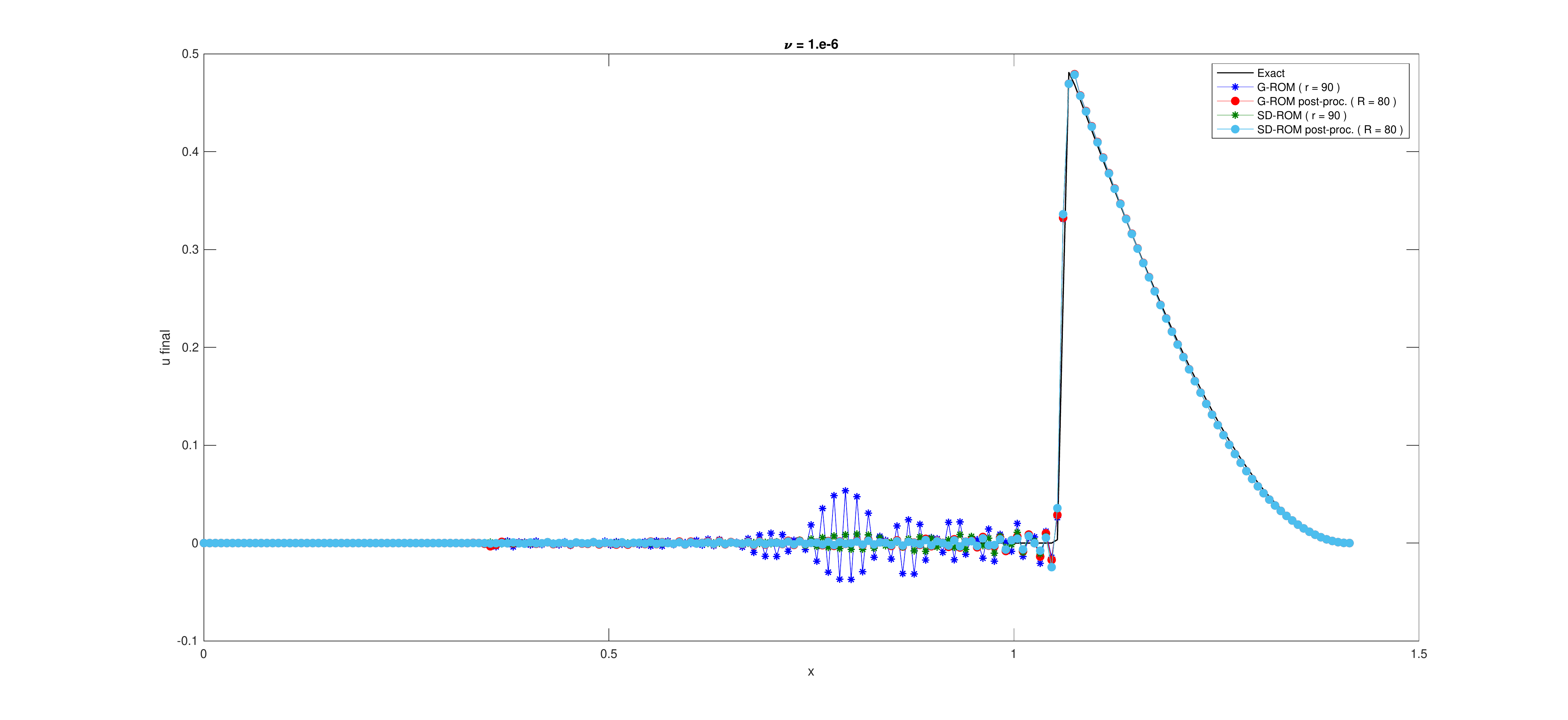}}
\caption{Example \ref{subsubsec:Ex1}: Final solution profiles along the mean diagonal for different ROM at $r=30, 60, 90$ (from top to bottom).}\label{fig:3}
\end{center}
\end{figure}

\begin{figure}[htb]
\begin{center}
\includegraphics[scale=0.125]{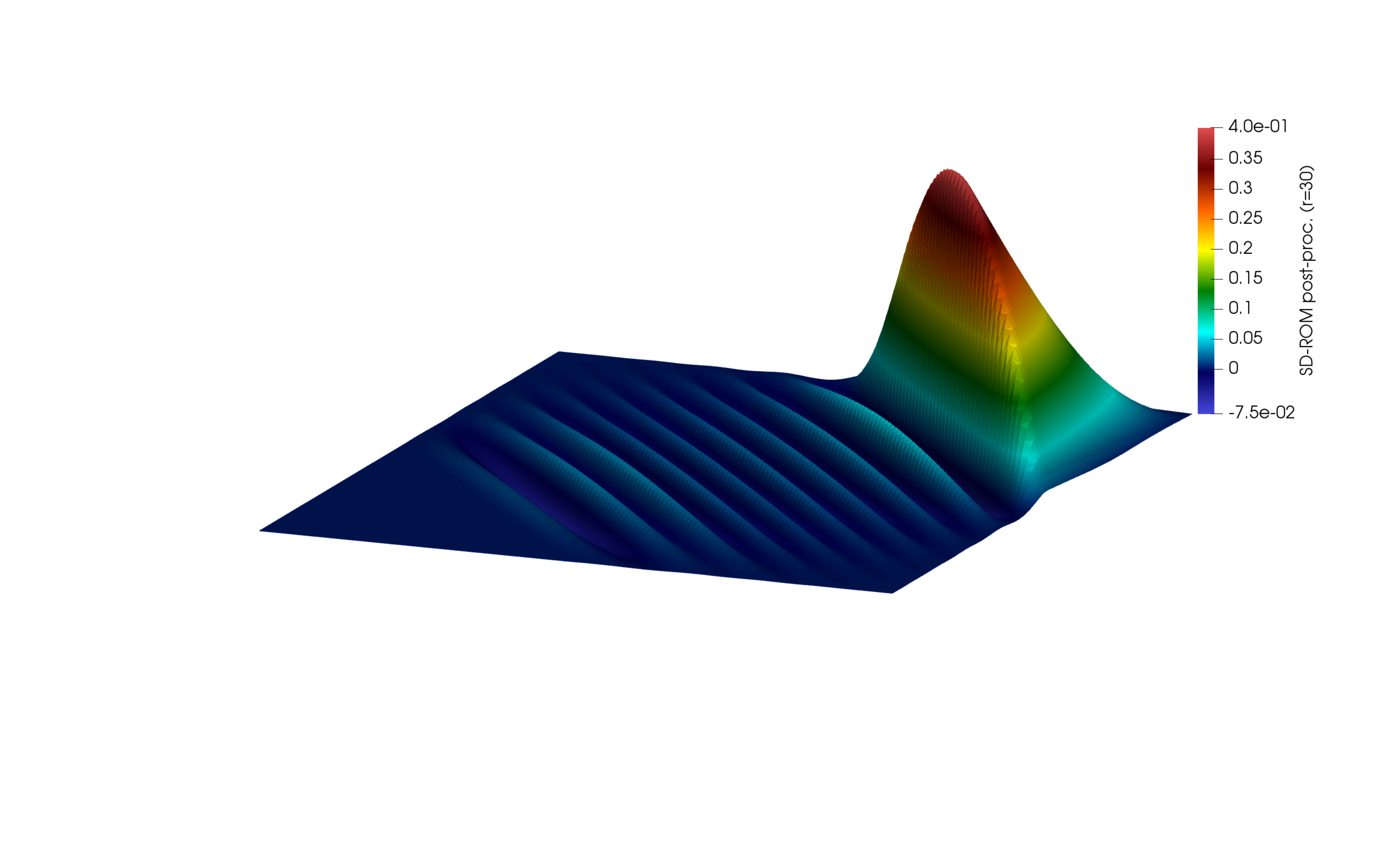}
\includegraphics[scale=0.125]{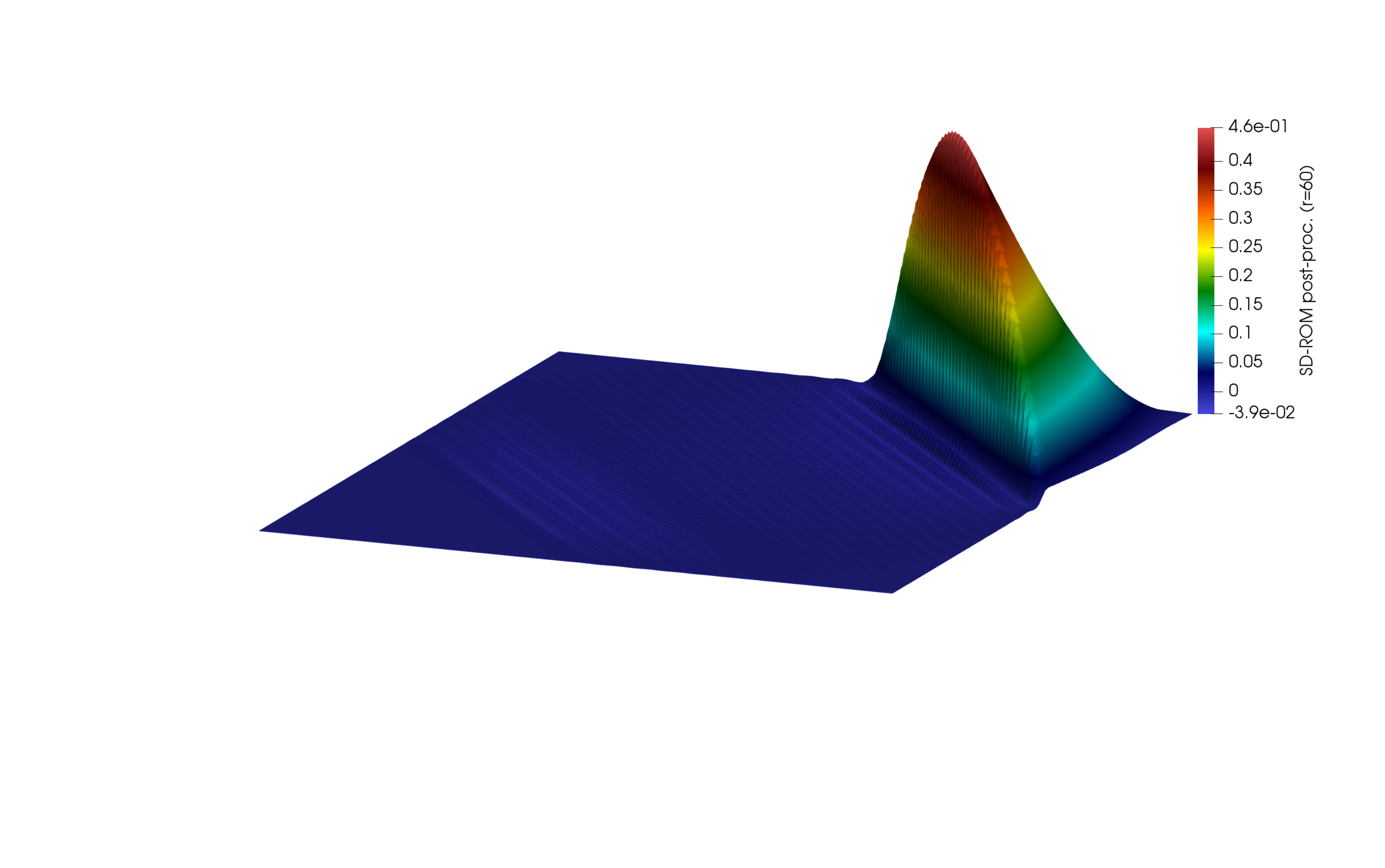}
\includegraphics[scale=0.125]{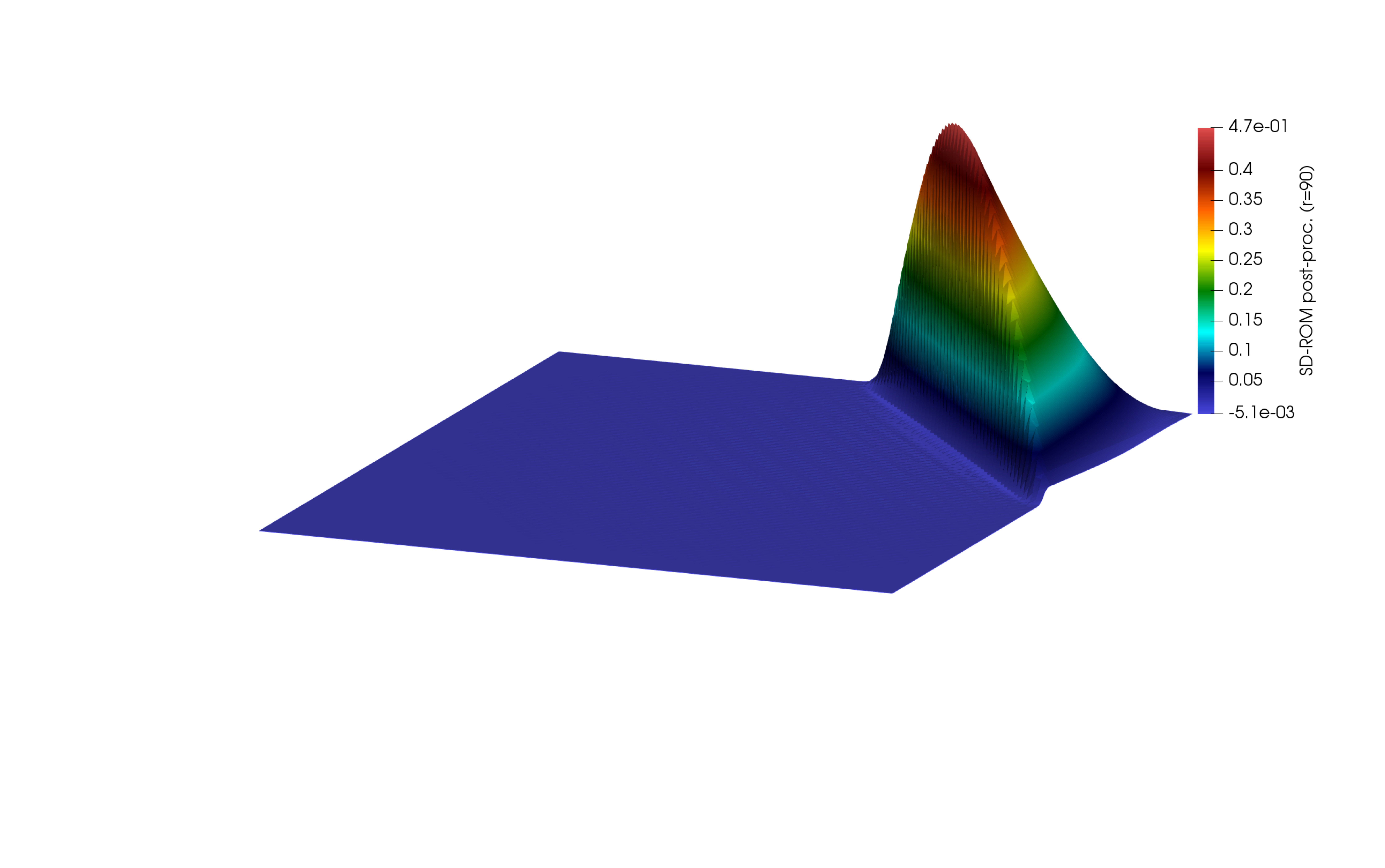}
\caption{Example \ref{subsubsec:Ex1}: Numerical solution for SD-ROM with online stabilizing post-processing at $T=1$ for $r=30, 60, 90$ (from top to bottom).}\label{fig:4}
\end{center}
\end{figure}

\begin{table}[htb]
$$\hspace{-0.1cm}
\begin{tabular}{|c|c|c|c|}
\hline
$\nu=10^{-6}$ & $r=30$ & $r=60$ & $r=90$\\
\hline
Captured system's $E_{kin} (\%)$ & $99.76$ & $99.98$ & $> 99.99$\\
\hline
\hline
$\nu=10^{-6}$ & \multicolumn{3}{|c|}{$e_{0}^{ROM}$}\\
\hline
Online methods & $r=30$ & $r=60$ & $r=90$\\
\hline
G-ROM & $0.3180$ & $0.1567$ & $0.1067$\\
\hline
G-ROM post-proc. & $0.3743$ & $0.1389$ & $0.0605$\\
\hline
SD-ROM & $0.2671$ & $0.1435$ & $0.0637$\\
\hline
SD-ROM post-proc. & $0.3465$ & $0.1383$ & $0.0579$\\
\hline
\end{tabular}$$\caption{Example \ref{subsubsec:Ex1}: Captured system's kinetic energy and $L^2$-norm of the deviation from the final exact solution profile along the mean diagonal for different ROM at $r=30, 60, 90$.}\label{tab:2}
\end{table}
\medskip
Figure \ref{fig:3} presents results for all considered ROM: the standard POD-Galerkin ROM \eqref{eq:POD-ROM} (G-ROM), the G-ROM with online stabilized post-processing (G-ROM post-proc.), the SD-POD-ROM \eqref{eq:SD-POD-ROM} (SD-ROM), and the SD-ROM with online stabilized post-processing (SD-ROM post-proc.). In particular, we show for the different methods the final solution profiles along the mean diagonal of the computational domain compared with the corres\-ponding exact solution profile, at $r=30, 60, 90$ (from top to bottom). One can observe that applying the online a-posteriori stabilization greatly improves results for the standard Galerkin-ROM (totally oscillatory), making it comparable with the stabilized SD-ROM, for which applying or not the online a-posteriori stabilization almost gives similar results. This is reflected by results depicted in table \ref{tab:2}, where the deviation $e_{0}^{ROM}$ from the final exact solution profile along the mean diagonal in a normalized discrete $L^{2}$-norm subject to \eqref{errROM} is displayed. One can see that, for $r=90$, SD-ROM post-proc. method almost reaches the same accuracy of the offline phase by almost suppressing the influence of noisy modes. Also, note that although the first $r=30$ POD modes already capture more than $99\%$ of the system's kinetic energy, all ROM yield poor quality results for which the peak of the front is not reached (the online stabilizing post-processing seems to be too numerical diffusive), and they display visible numerical oscillations, reflecting the complexity of the problem. Augmenting the number of POD modes allows to reach the peak of the front for all methods. However, whereas the solution of the G-ROM remains globally polluted with spurious oscillations, the application to it of the online a posteriori stabilization already reduces to few oscillations and localize them mainly near the steep layer, allowing to compute a rather accurate solution in this case, comparable with the one of the stabilized SD-ROM and of the offline phase. In figure \ref{fig:4}, we show the numerical solution at $T=1$ for the best performing SD-ROM with online a posteriori stabilization for $r=30, 60, 90$ (from top to bottom). With this method, numerical unphysical oscillations are practically eliminated by gradually increasing the number of POD modes.

\subsubsection{Case $\nu = 10^{-8}$}\label{subsubsec:Ex2}
\begin{figure}[htb]
\begin{center}
\centerline{\includegraphics[scale=0.4]{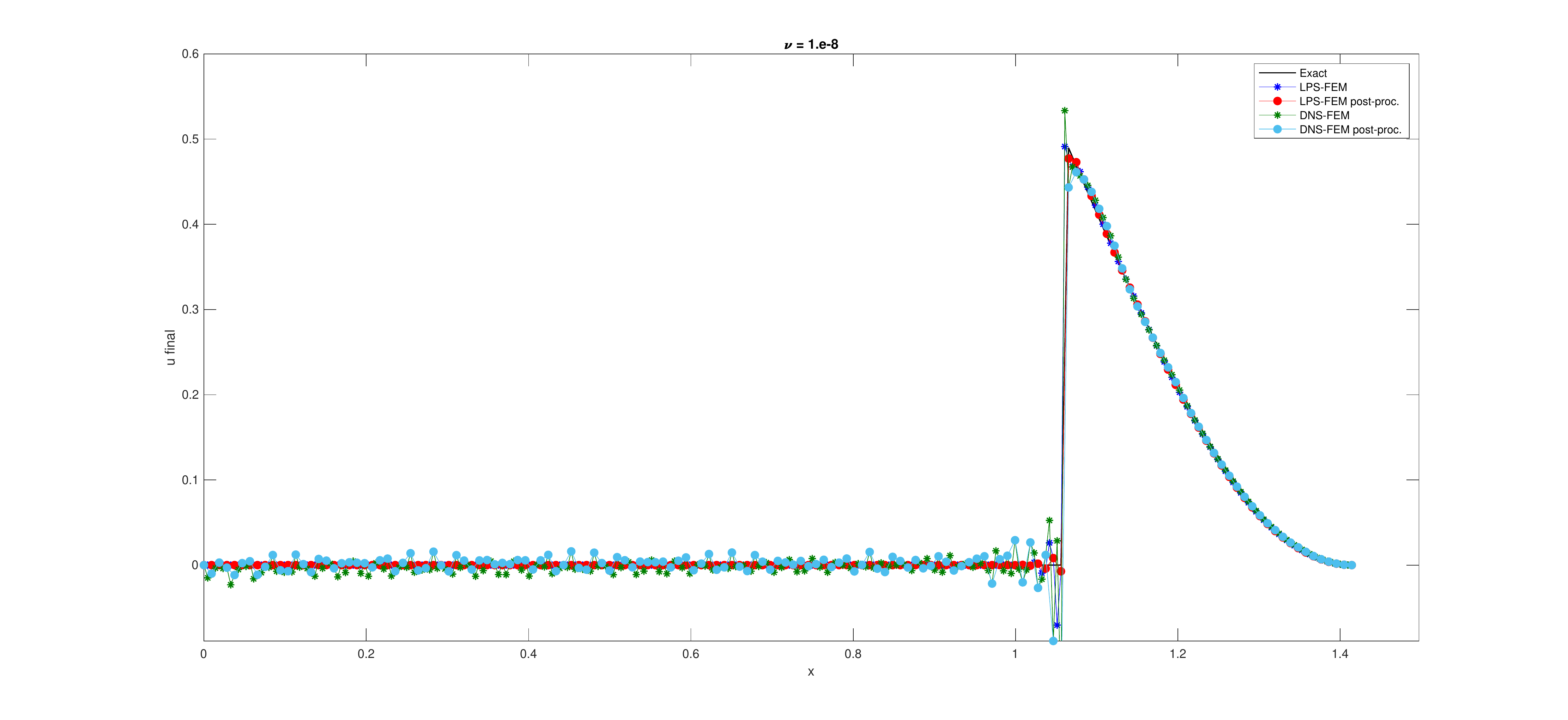}}
\caption{Example \ref{subsubsec:Ex2}: Final solution profiles along the mean diagonal for different FOM.}\label{fig:5}
\end{center}
\end{figure}

In this case, we consider a uniform triangular mesh with mesh size $h=9.43\cdot 10^{-3}$. Thus, a finer grid with respect to the previous case is used, which is necessary to maintain numerical diffusion within reasonable limits. Nevertheless, it remains relatively coarse with respect to the width of the internal layer. Again, we tested different FOM: DNS-FEM, DNS-FEM post-proc., LPS-FEM, and LPS-FEM post-proc. In figure \ref{fig:5}, we show for the different methods the final solution profiles along the mean diagonal of the computational domain compared with the corresponding exact solution profile.

\begin{figure}[htb]
\begin{center}
\centerline{\includegraphics[scale=0.4]{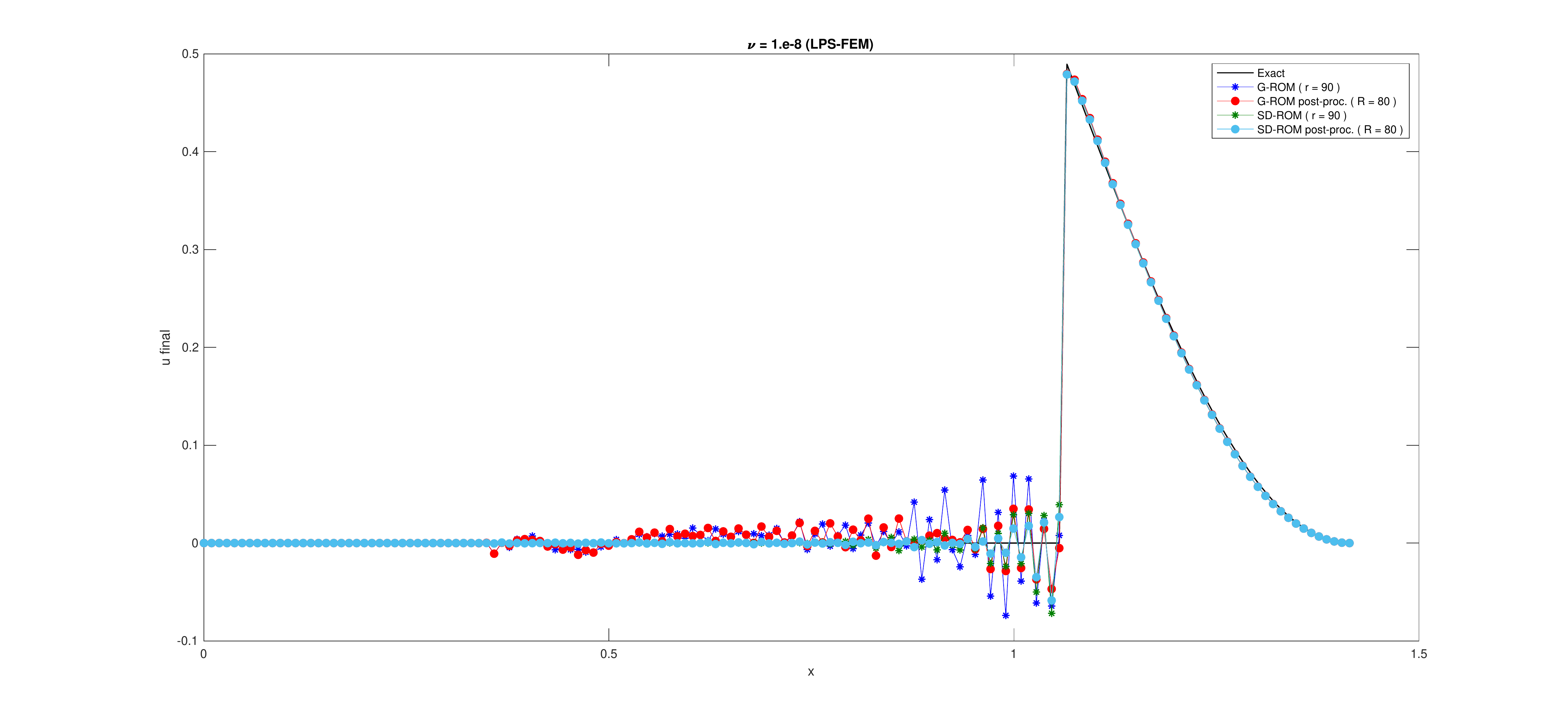}}
\caption{Example \ref{subsubsec:Ex2}: Final solution profiles along the mean diagonal for different ROM at $r=90$ using noisy POD data from LPS-FEM.}\label{fig:6}
\end{center}
\end{figure}

\begin{table}[htb]
$$\hspace{-0.1cm}
\begin{tabular}{|c|c|}
\hline
Offline methods & $e_{0}^{FOM}$, $\nu=10^{-8}$\\
\hline
DNS-FEM & $0.1816$\\
\hline
DNS-FEM post-proc. & $0.1345$\\
\hline
LPS-FEM & $0.1247$\\
\hline
LPS-FEM post-proc. & $0.0393$\\
\hline
\end{tabular}$$\caption{Example \ref{subsubsec:Ex2}: $L^2$-norm of the deviation from the final exact solution profile along the mean diagonal for different FOM.}\label{tab:3}
\end{table}
\medskip
Offline results proves again the necessity to consider LPS method to avoid globally spurious oscillations, but also that the application of the a-posteriori stabilization greatly improves the results of the LPS-FEM in this case. Indeed, error levels decrease from $12\%$ to $4\%$ when applying stabilizing post-processing to LPS-FEM, as shown in table \ref{tab:3}. Also, if we proceed by constructing POD basis from LPS-FEM (without stabilizing post-processing), being more influenced by spurious oscillations, it leads to online numerical solutions that are globally polluted with high spurious oscillations even for $r=90$, whatever it is the applied ROM, as shown in figure \ref{fig:6}.

\begin{figure}[htb]
\begin{center}
\centerline{\includegraphics[scale=0.4]{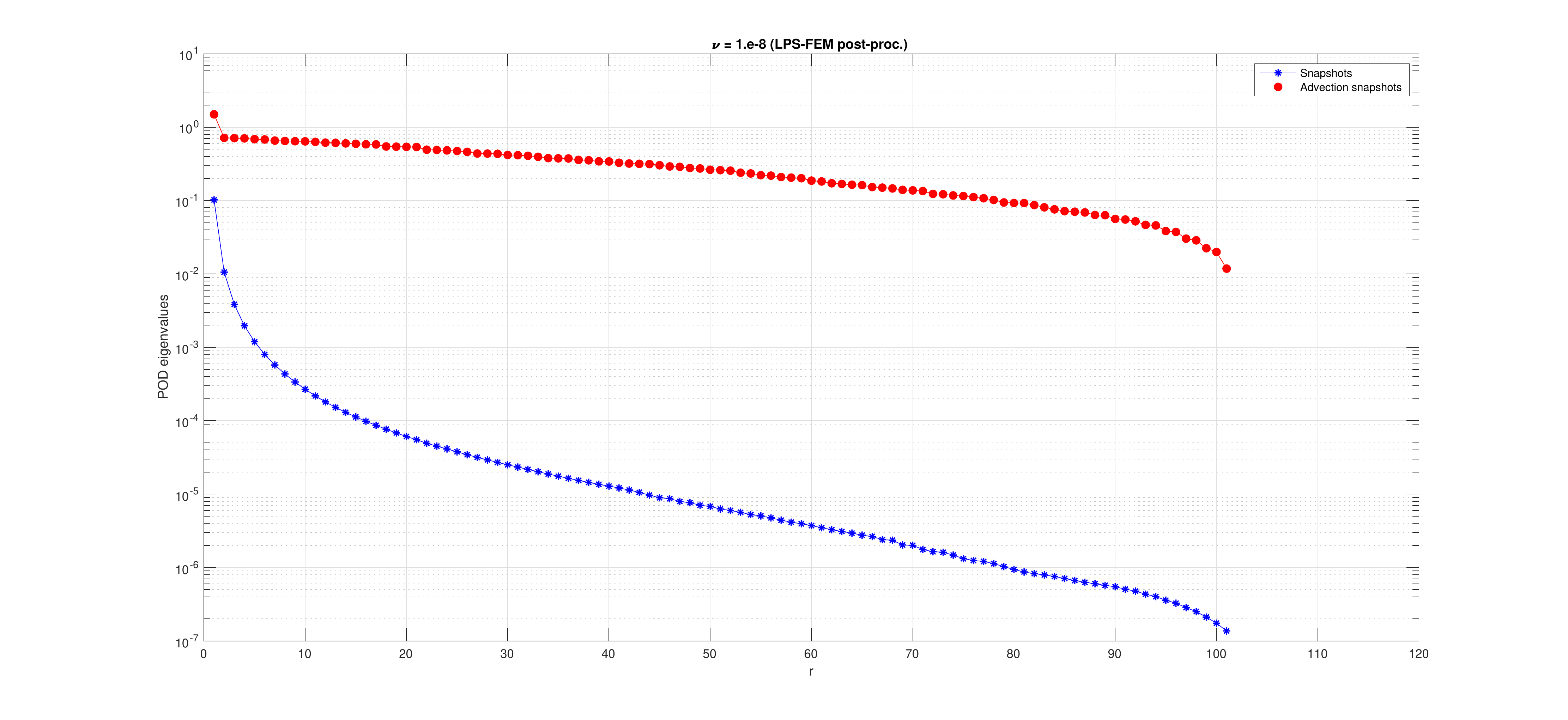}}
\caption{Example \ref{subsubsec:Ex2}: POD eigenvalues.}\label{fig:7}
\end{center}
\end{figure}

\medskip
Thus, we decided to proceed by constructing POD basis by using LPS-FEM with stabilizing post-processing, to limit the influence of POD noisy data in the online phase. In figure \ref{fig:7}, we show the decay of POD eigenvalues associated both to the snapshots correlation matrix \eqref{CorrM} and the advection correlation matrix \eqref{eq:Khat} in this case. Again, one can observe that the decay of the POD eigenvalues associated to the advection correlation matrix is rather slow, due to the very low diffusion. However, adding the corresponding stabilization term in the online phase greatly improves the results over the standard POD-ROM also in this case.

\begin{figure}[htb]
\begin{center}
\centerline{\includegraphics[scale=0.35]{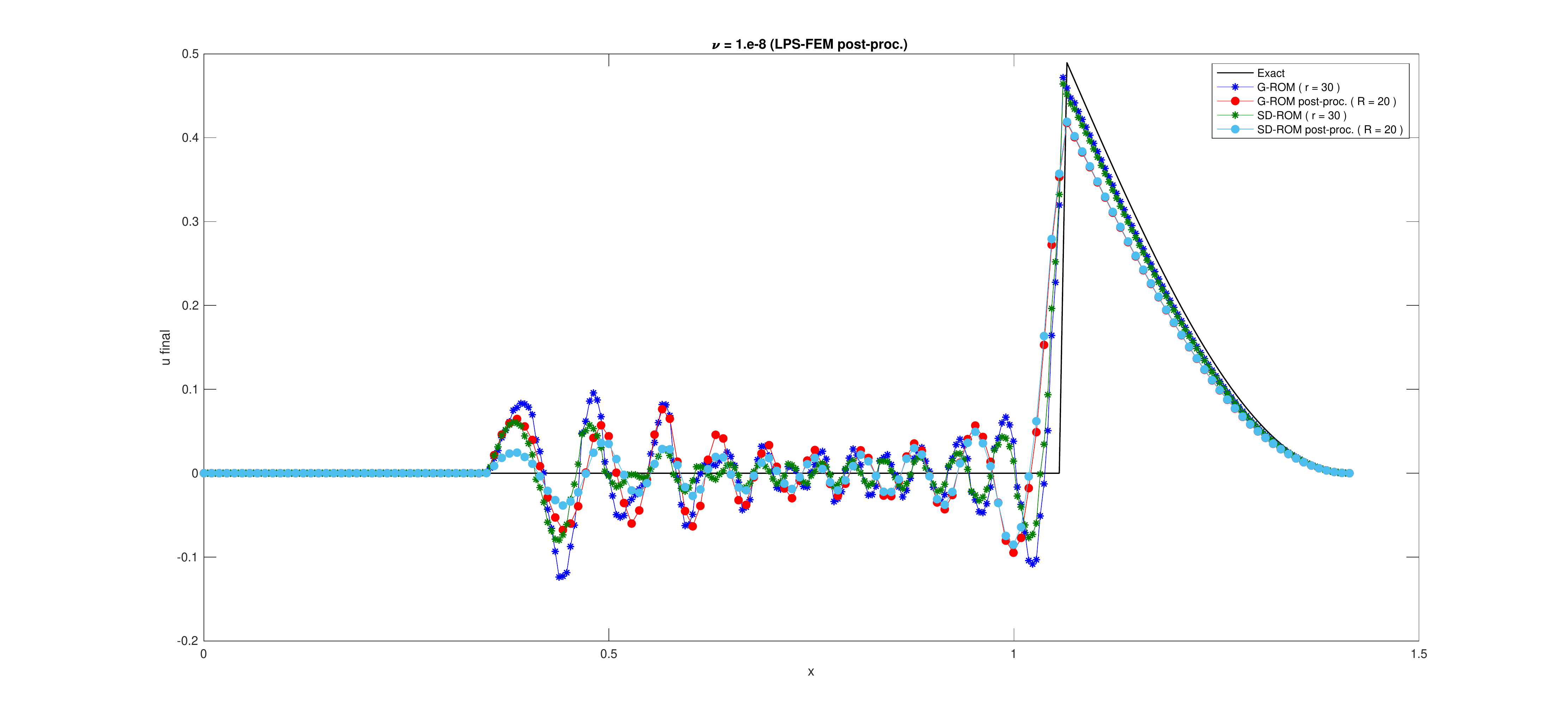}}
\centerline{\includegraphics[scale=0.35]{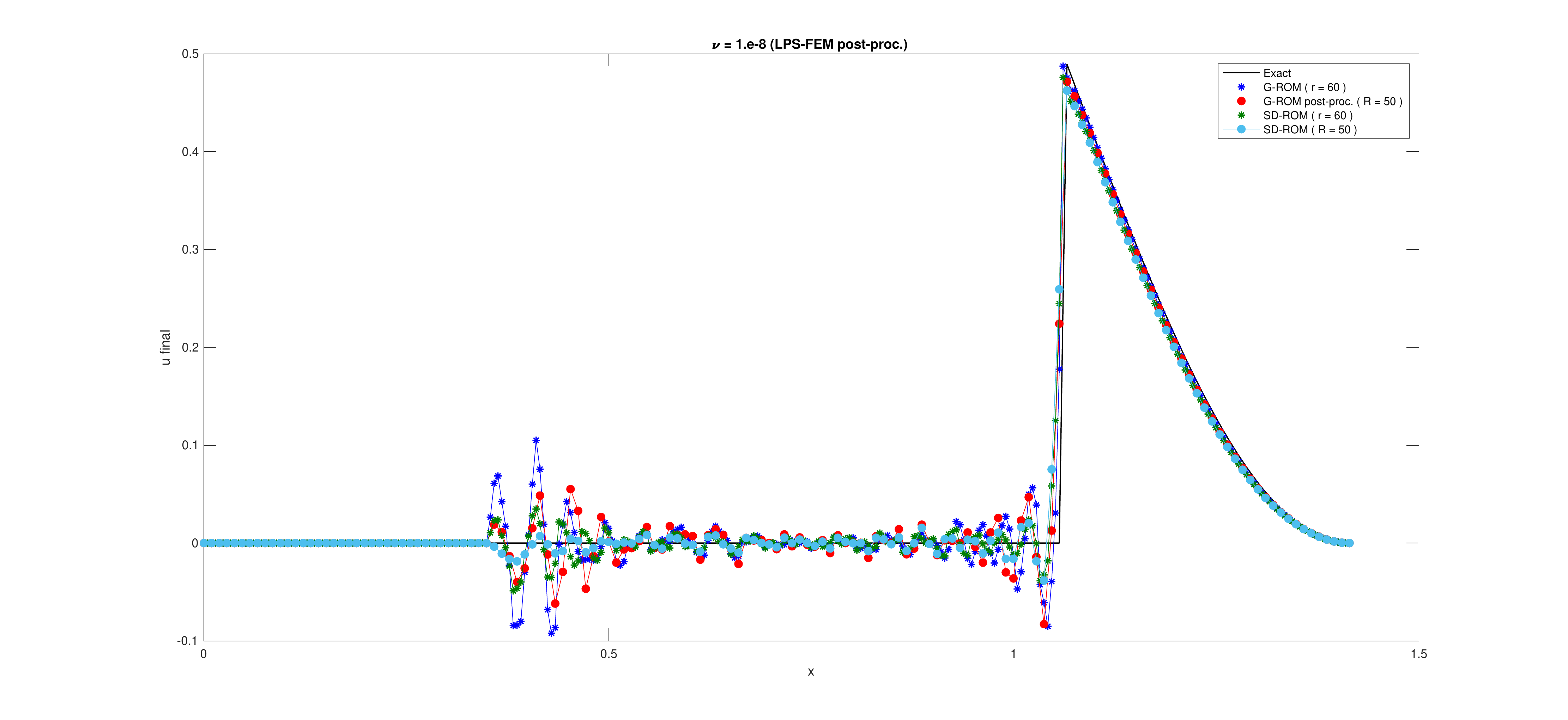}}
\centerline{\includegraphics[scale=0.35]{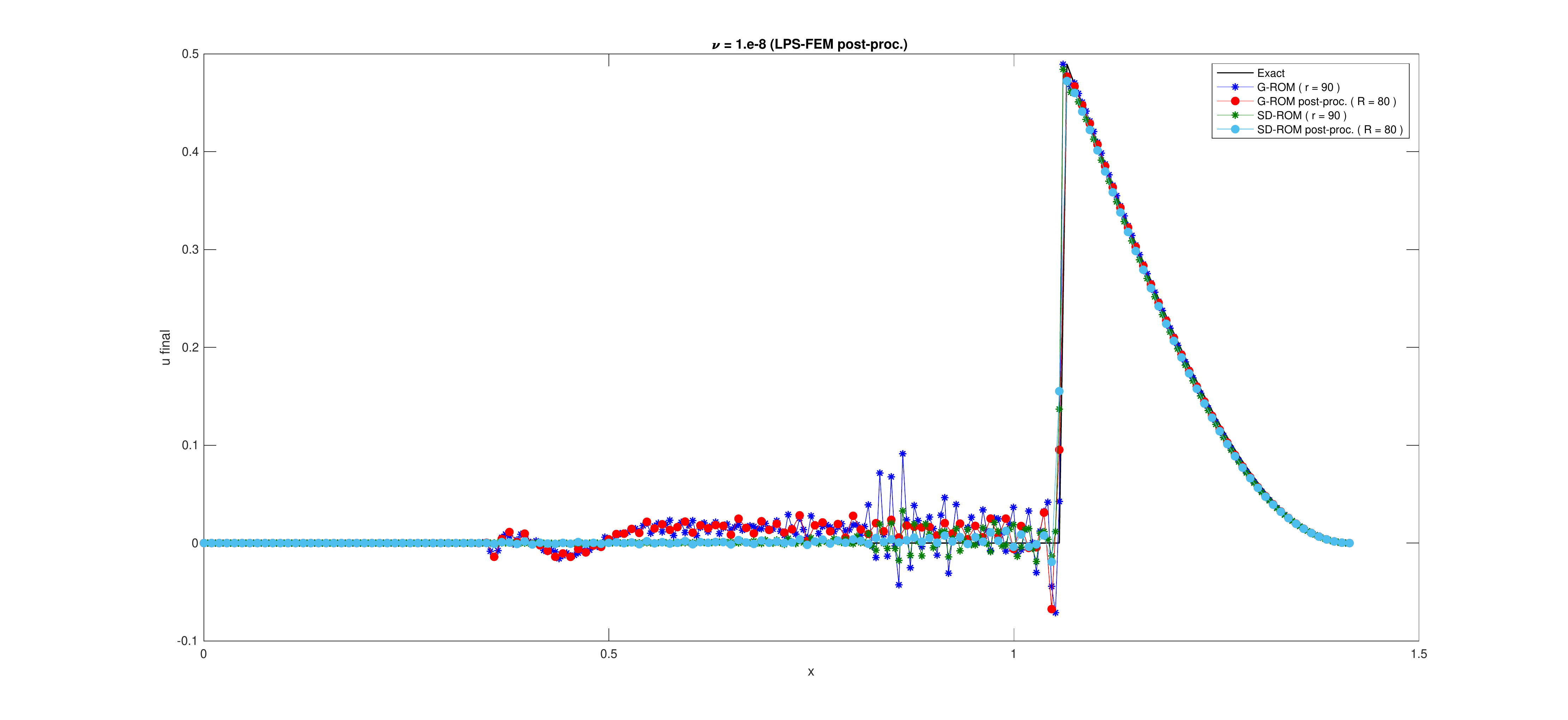}}
\caption{Example \ref{subsubsec:Ex2}: Final solution profiles along the mean diagonal for different ROM at $r=30, 60, 90$ (from top to bottom).}\label{fig:8}
\end{center}
\end{figure}

\begin{figure}[htb]
\begin{center}
\includegraphics[scale=0.125]{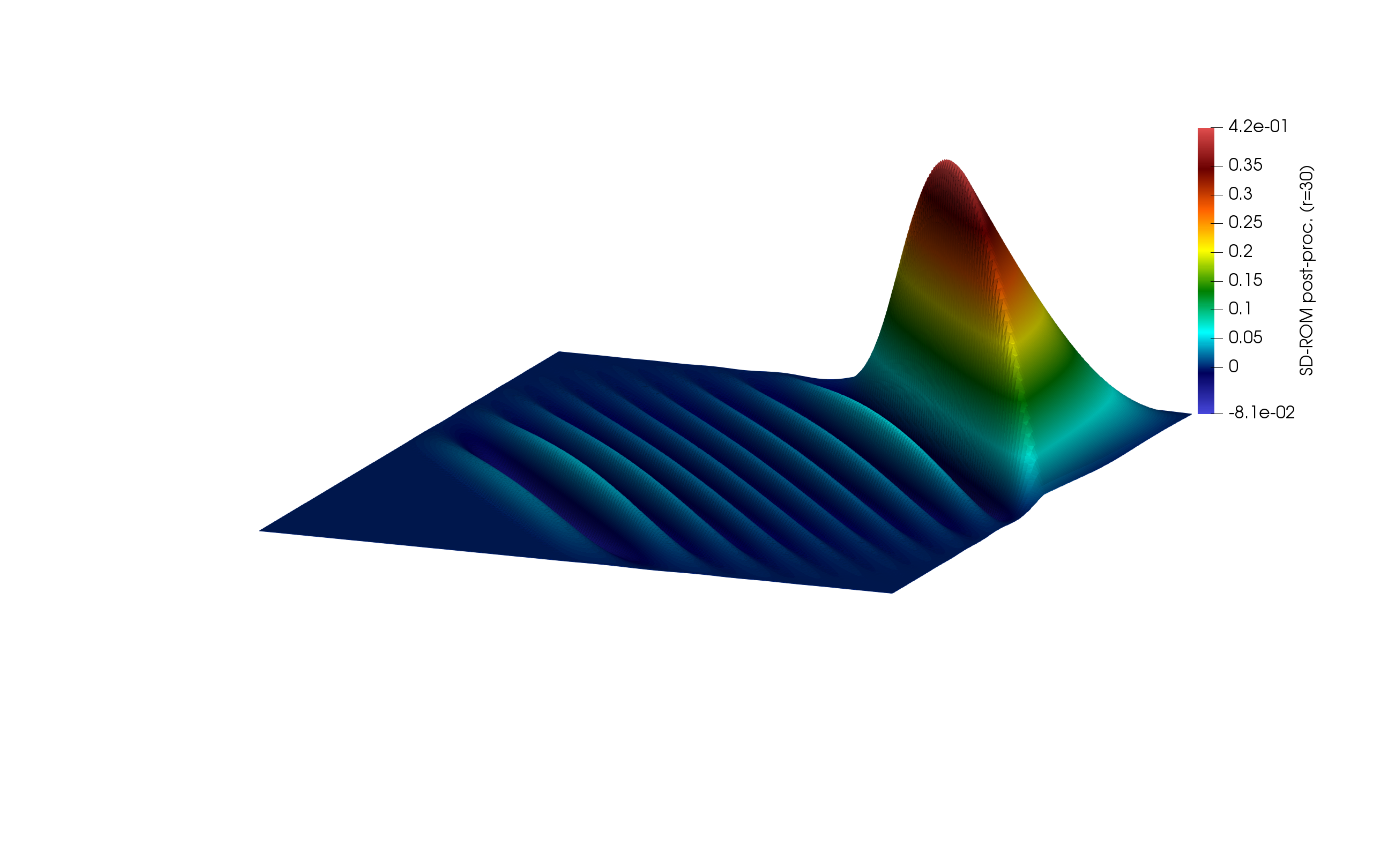}
\includegraphics[scale=0.125]{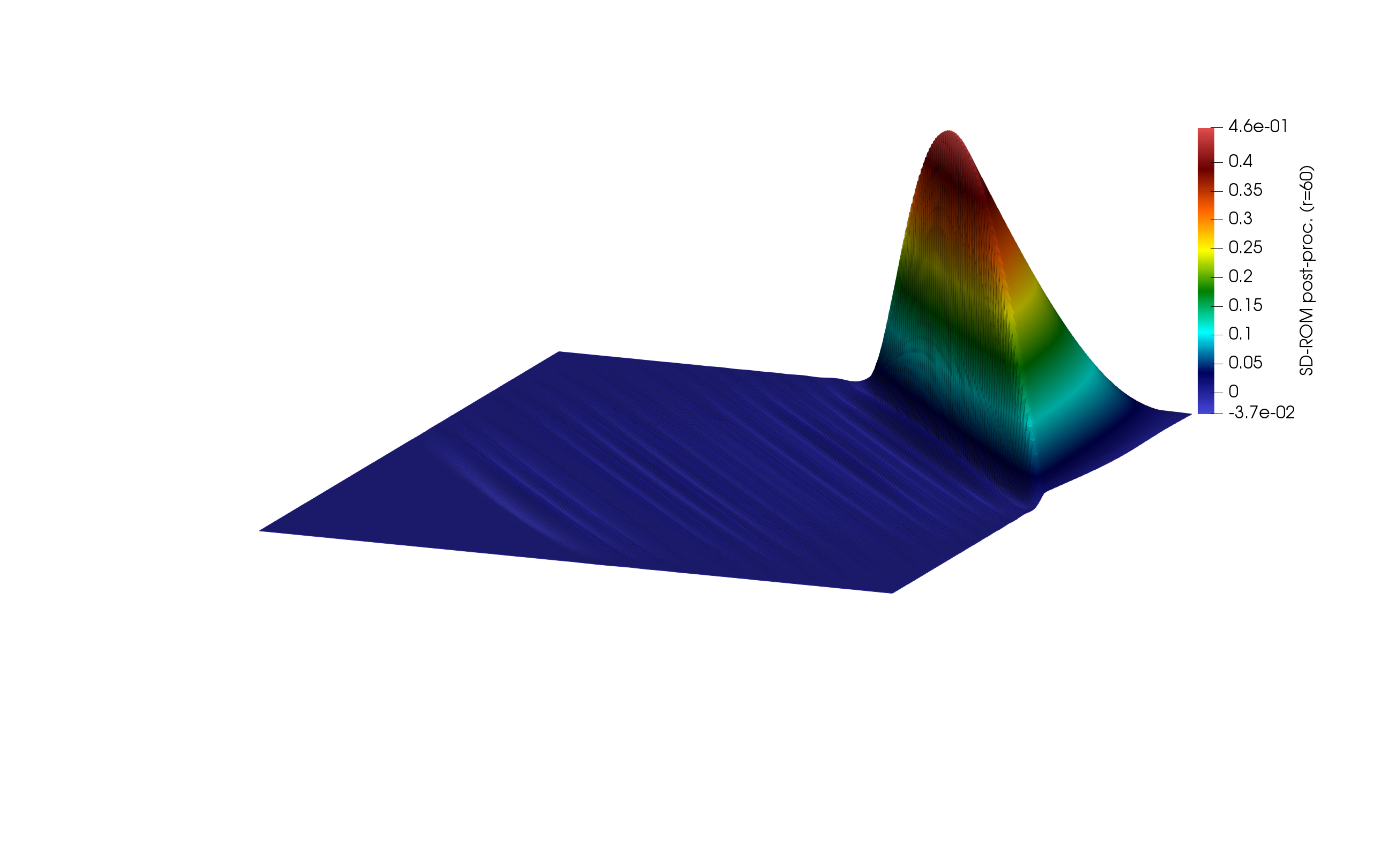}
\includegraphics[scale=0.125]{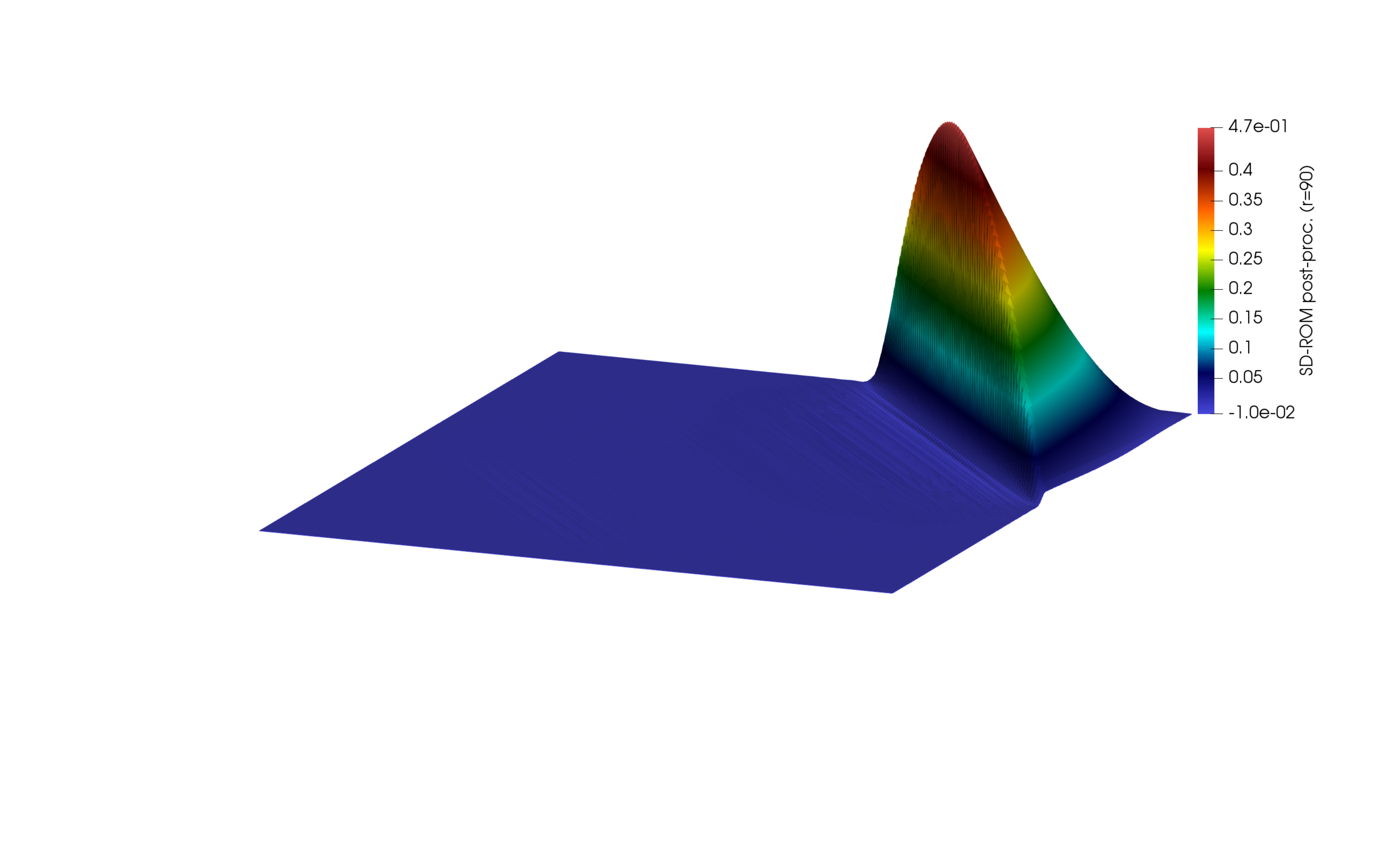}
\caption{Example \ref{subsubsec:Ex2}: Numerical solution for SD-ROM with online stabilizing post-processing at $T=1$ for $r=30, 60, 90$ (from top to bottom).}\label{fig:9}
\end{center}
\end{figure}

\medskip
Figure \ref{fig:8} presents results for all considered ROM: G-ROM, G-ROM post-proc., SD-ROM, and SD-ROM post-proc. One can observe that results for G-ROM (with and without online 
a-posteriori stabilization) are globally quite oscillatory, even at $r=90$. However, applying SD-ROM already localizes oscillations just near the moving steep layer, and also 
SD-ROM with online stabilizing post-processing allows to further improve results, maintaining the amplitude of oscillations in a reasonable low range. This is reflected by results depicted in table \ref{tab:4}. One can see that, for $r=90$, SD-ROM post-proc. method approaches the accuracy of the offline phase by considerably suppressing the influence of noisy modes. Again, note that although the first $r=30$ POD modes already capture more than $99\%$ of the system's kinetic energy, all ROM yield poor quality results for which the peak of the front is not reached (the online stabilizing post-processing seems to be too numerical diffusive), and they display globally spread numerical oscillations, reflecting the extreme complexity of the problem. Augmenting the number of POD modes allows to reach the peak of the front for all methods. However, whereas the solution of the G-ROM (with and without online a-posteriori stabilization in this case) remains globally polluted with spurious oscillations, the SD-ROM notably reduces the amplitude of oscillations, 
and its combination with online stabilizing post-processing allows to compute a rather accurate solution in this case, comparable with the one of the offline phase. In figure \ref{fig:9}, we show the numerical solution at $T=1$ for the best performing SD-ROM with online a posteriori stabilization for $r=30, 60, 90$ (from top to bottom). Again, with this method, numerical unphysical oscillations are practically eliminated by gradually increasing the number of POD modes.

\begin{table}[htb]
$$\hspace{-0.1cm}
\begin{tabular}{|c|c|c|c|}
\hline
$\nu=10^{-8}$ & $r=30$ & $r=60$ & $r=90$\\
\hline
Captured system's $E_{kin} (\%)$ & $99.71$ & $99.96$ & $> 99.99$\\
\hline
\hline
$\nu=10^{-8}$ & \multicolumn{3}{|c|}{$e_{0}^{ROM}$}\\
\hline
Online methods & $r=30$ & $r=60$ & $r=90$\\
\hline
G-ROM & $0.3086$ & $0.1676$ & $0.1224$\\
\hline
G-ROM post-proc. & $0.3733$ & $0.1493$ & $0.0884$\\
\hline
SD-ROM & $0.2596$ & $0.1463$ & $0.0675$\\
\hline
SD-ROM post-proc. & $0.3417$ & $0.1449$ & $0.0589$\\
\hline
\end{tabular}$$\caption{Example \ref{subsubsec:Ex2}: Captured system's kinetic energy and $L^2$-norm of the deviation from the final exact solution profile along the mean diagonal for different ROM at $r=30, 60, 90$.}\label{tab:4}
\end{table}

\section{Summary and conclusions}\label{sec:Concl}
In this work, we have proposed to improve the stabilized POD-ROM introduced by S. Rubino in \cite{Rubino18} to deal with the numerical simulation of advection-dominated advection-diffusion-reaction equations. In particular, we introduced a stabilizing post-processing strategy that has proved to be very useful when considering very low diffusion coefficients, i.e. in the strongly advection-dominated regime. This strategy has been applied both for the offline phase, to produce less noisy snapshots and, as consequence, limit the influence of POD noisy modes in the online phase, and the reduced order method to compute more stable and accurate online solutions. The new process of a posteriori stabilization has been detailed in a general framework and applied to advection-diffusion-reaction problems. The performed numerical studies have shown the potential of the new ROM in handling strongly advection-dominated cases, also tested for long time integrations on periodic systems, by extremely limiting spurious oscillations and thus obtaining rather accurate results in this framework. To remove the few remaining oscillations, one could think to apply more complex shock or discontinuity capturing methods (see \cite{JB07} for a detailed review) and try to adapt them to the POD-ROM framework as future interesting research topic. Also, one could carry out a similar numerical investigation of the significantly more challenging Navier--Stokes equations in view of computing more complex convection-dominated and turbulent flows.

\medskip

{\sl Acknowledgments:} This work has been partially supported by the Spanish Government - Feder EU grant MTM2015-64577-C2-1-R. The first author has received financial support from the French State in the frame of the "Investments for the future IdEx Bordeaux (Initiative d'Excellence de l'Universit\'e de Bordeaux) Programme, reference ANR-10-IDEX-03-02. The third author would gratefully acknowledge the financial support received from IdEx Bordeaux International Post-Doc Programme during his initial postdoctoral research involved in this article. The research of the third author has been also funded by the Spanish State Research Agency through the national programme €œJuan de la Cierva-incorporaci\'on 2017.

\bibliographystyle{abbrv}
\bibliography{Biblio_APS-POD-ROM}


\end{document}